\documentclass[%
 reprint,
 superscriptaddress,
nofootinbib,
amsmath,amssymb,
aps,
prd,
]{revtex4-2}

\usepackage{graphicx}
\usepackage{dcolumn}
\usepackage{natbib}
\usepackage{amsfonts,amsbsy}
\usepackage[utf8]{inputenc}
\usepackage{placeins}
\usepackage{xcolor}
\usepackage[colorlinks]{hyperref}

\newcommand{\Msun}{M_{\odot}}
\newcommand{\fm}{\,\mathrm{fm}}
\newcommand{\MeV}{\,\mathrm{MeV}}
\graphicspath{{}}

\newif\ifreviewmarks
\reviewmarkstrue
\ifreviewmarks
       
  \newcommand{\revnote}[1]{{\color{red}#1}}   
       
  \definecolor{revorange}{RGB}{204,85,0}
   
  \definecolor{revbrown}{RGB}{121,68,21}
    
  \definecolor{revgreen}{RGB}{0,110,60}
  \newcommand{\revg}[1]{{\color{revgreen}#1}}

  \hypersetup{colorlinks=true, citecolor=blue, linkcolor=red, urlcolor=teal}
  \usepackage{fancyhdr}
\else
  
  \newcommand{\revnote}[1]{}

  \newcommand{\revg}[1]{#1}
  \hypersetup{hidelinks}
\fi

\begin{document}

\title{Universal relations for hybrid stars and their quark cores: 
a systematic scan of the quark-matter parameter space}

\author{J.\,P.~Carlomagno}
\email{carlomagno@fisica.unlp.edu.ar}
\affiliation{IFLP, CONICET -- Departamento de F\'{\i}sica, Facultad de Ciencias Exactas,
             Universidad Nacional de La Plata, C.C.~67, 1900 La Plata, Argentina}

\author{G.\,A.~Contrera}
\email{contrera@fisica.unlp.edu.ar}
\affiliation{IFLP, CONICET -- Departamento de F\'{\i}sica, Facultad de Ciencias Exactas,
             Universidad Nacional de La Plata, C.C.~67, 1900 La Plata, Argentina}

\author{A.\,G.~Grunfeld}
\email{agrunfeld@iafe.uba.ar}
\affiliation{CONICET, Godoy Cruz 2290, Buenos Aires, Argentina}
\affiliation{Instituto de Astronom\'{\i}a y F\'{\i}sica del Espacio (IAFE, CONICET-UBA),
             Ciudad Universitaria, 1428, Buenos Aires, Argentina}

\author{G.~Lugones}
\email{german.lugones@ufabc.edu.br}
\affiliation{Universidade Federal do ABC, Centro de Ci\^encias Naturais e Humanas,
             Avenida dos Estados 5001 -- Bang\'u, CEP 09210-580, Santo Andr\'e, SP, Brazil}

% ─────────────────────────────────────────────────────────────

\begin{abstract}
Universal, equation-of-state (EOS) insensitive relations among the global
properties of compact stars are well established for hadronic neutron stars
and for self-bound strange stars. We ask whether analogous relations hold
for the quark core of a hybrid star, and whether the core can be
constrained from the surface. Describing quark matter with the three-flavor MIT bag
model with repulsive vector interactions, matched to the hadronic DD2 EOS
through a Maxwell construction, we scan the full $(B^{1/4},G_V)$ plane and
obtain 1112 hybrid EOSs with maximum masses above $2\,M_\odot$,
broadly compatible with current multimessenger constraints.
For every stable configuration we evaluate the mass, radius, moment of
inertia, and tidal deformability both at the stellar surface and at the
boundary of the quark core. 
Global quantities follow the relations of
hadronic stars---the $I$--Love relation to within $0.5\%$---while core
quantities follow those of self-bound strange stars with median
deviations of $6$--$9\%$. 
We explain the latter by expanding the structure of the core in powers of
its compactness: to first order, the core mass, baryon mass, moment of
inertia, and tidal deformability depend only on the compactness and on the
properties of the hadron--quark interface, chiefly the ratio
$\xi=p_{\rm tr}/\varepsilon_+$ of transition pressure to quark energy
density. The strange-star relations are the $\xi\to0$ member of this
family, and the deviations from them measure $\xi$. The fractional core
mass and radius of all configurations follow the normalized mass profile
of the star, and no global observable allows the core size to be inferred.
Evanescent cores dominate the viable parameter space, while sizable cores
require low bag constants and strong vector couplings.
\end{abstract}

\maketitle

\section{Introduction}
\label{sec:intro}
%==============================================================

Whether deconfined quark matter exists in the cores of compact stars is a
question that sits at the intersection of QCD, nuclear physics, and
gravitational theory. If a first-order hadron--quark transition occurs at
the densities reached inside neutron stars, the resulting hybrid
configurations---a quark core wrapped in a hadronic envelope---would
differ structurally from their purely nucleonic counterparts, yet the
observational signatures of this difference remain
elusive~\cite{Alford:2004pf,Zdunik:2012dj,Baym:2017whm}. The difficulty
is not a lack of data: tidal deformabilities from
GW170817~\cite{TheLIGOScientific:2017qsa,Abbott:2018exr}, mass--radius
measurements from
NICER~\cite{Riley:2019yda,Miller:2019cac,Riley:2021pdl,Miller:2021qha},
and pulsar masses above
$2\,M_\odot$~\cite{Demorest:2010bx,Antoniadis:2013pzd,Fonseca:2021wxt}
now constrain the equation of state with considerable precision.
The difficulty is that these observables are largely degenerate: very
different interior compositions can produce nearly identical global
properties.

That hybrid stars may be observationally indistinguishable from
purely hadronic ones, their mass--radius relations overlapping over
the entire observed range, was recognized two decades ago, with a simple
bag-model description of the quark phase, and is known as the ``masquerade''
problem~\cite{Alford:2004pf}. Less explored is the variety of interior
configurations that can hide behind this common exterior. A hybrid star may
masquerade because its quark core is tiny and appears only at the very end
of the stable sequence, or because a sizable core, through the combination
of the energy-density jump and the stiffness of the quark phase, happens to
leave the global observables close to their hadronic values. Which of these
situations prevails, how the transition pressure, the density jump, and the
resulting core size combine to produce it, and how much of the admissible
parameter space of a microscopic model is occupied by each, are questions
that only a systematic exploration of that parameter space can answer. At
the same time, and although the masquerade would seem to suggest otherwise,
it is worth examining explicitly whether relations exist that connect what
is measured at the stellar surface to the properties of the core or,
failing that, whether universal relations hold among the properties of the
core itself, as they do among the global properties of hadronic and strange
stars.

Universal relations offer a useful framework for approaching both questions. 
The EOS-insensitive $I$--Love--$Q$
correlations~\cite{Yagi:2013bca,Yagi:2013awa}, later extended to rapid
rotation~\cite{Doneva:2013rha,Pappas:2013naa,Chakrabarti:2013tca},
exotic compositions~\cite{Lenka:2018ehb,Raduta:2020fdn,Yagi:2015hda,%
Pretel:2024pem}, and self-bound strange quark
stars~\cite{Urbanec:2013fs,Lugones:2025lis}, have been used to
estimate quantities that are difficult to measure directly, to break
parameter degeneracies in gravitational-wave analysis, and to test
gravity without assuming a specific EOS~\cite{Yagi:2016bkt}. For
hybrid stars they hold at the few-percent level despite the density
discontinuity introduced by the phase
transition~\cite{Paschalidis:2017qmb}, also under rapid
rotation~\cite{Roy:2025rdy}; known departures are restricted to
specific settings, such as elastic effects in crystalline
color-superconducting cores~\cite{Dong:2024opz} and oscillation-mode
relations in the slow-conversion
scenario~\cite{Ranea-Sandoval:2022izm}.
These analyses, however, concern quantities defined at the stellar
surface only; the quark core, read at its own boundary, has not been
subjected to the same test.

In this paper we therefore address both the global and core-level universality questions within a single microscopic framework, exploring its parameter space in depth.
We describe the quark phase with the
three-flavor MIT bag model supplemented by repulsive vector
interactions~\cite{Franzon:2016urz,Lopes2021_I,Lopes2021_II}, match it
to the hadronic DD2 EOS~\cite{Typel:2009sy} through a Maxwell
construction, and scan the plane spanned by the bag constant $B^{1/4}$
and the vector coupling $G_V$, obtaining $1112$ hybrid EOSs with
maximum masses above $2\,M_\odot$, broadly compatible with current
multimessenger constraints. The scan is exhaustive on the quark side; the hadronic
side is kept fixed, a choice whose implications we assess in the
concluding section. For every stable configuration of every sequence we
compute the gravitational and baryon masses, radius, moment of inertia,
and tidal deformability \emph{twice}: at the stellar surface and at the
boundary of the quark core. 
Comparing the two evaluations of each configuration allows us to test
universality at both levels---and thereby to address both questions
across the entire viable parameter space. We find that universality holds at both levels, but for
different reasons, and at the core level we trace it to an analytic origin:
an expansion of the core structure in powers of its compactness, in which
the quark EOS enters only through the properties of the interface.

A complementary question concerns the geometry of the quark core itself.
While the universal relations discussed above are formulated in terms of
quantities defined at a surface, the internal mass profile $m(r)$ contains
additional information about how the deconfined core is embedded within the
star. The relation between the mass and radius enclosed by the phase
boundary is a direct probe of that profile: if $m(r)/M$ were nearly the same
function of $r/R$ for all configurations, the fractional core mass would be
determined by the fractional core radius alone, independently of the quark
EOS and of the position of the star along its sequence. We show in
Sec.~\ref{sec:core_characterization} that this is the case to within a few
percent of the stellar mass, so that the relation is a property of the
normalized structure of the star rather than of the core, and we ask there
whether such internal relations can be inverted---whether the core can be
inferred from global observables.

The paper is organized as follows. Section~\ref{sec:formalism} summarizes
the hybrid EOS, the structure equations, the notation, and the numerical
method. Section~\ref{sec:results} presents the parameter-space exploration,
the mass--radius sequences, and the classification of the viable EOSs into
evanescent and extended cores. Section~\ref{sec:universal} tests the
universal relations for star and core and presents the small-core expansion
that explains the core relations, whose derivation is given in
Appendix~\ref{app:smallcore}. Section~\ref{sec:core_characterization}
analyzes the normalized mass profile and the core-mass--core-radius
relation, and asks whether the core can be inferred from global
observables. Section~\ref{sec:conclusions} contains our conclusions. We use
geometrized units with $G=c=1$ unless otherwise stated.

%==============================================================
\section{Formalism}
\label{sec:formalism}
%==============================================================

%--------------------------------------------------------------
\subsection{Hybrid equation of state}
\label{sec:EOS}
%--------------------------------------------------------------

\label{sec:hadronic}%
The hadronic phase is described by a unified crust--core EOS for cold,
neutrino-free, $\beta$-equilibrated, charge-neutral $npe\mu$ matter, based on
the density-dependent relativistic mean-field DD2
parametrization~\cite{Typel:2009sy}: the outer crust follows the BSk22
functional~\cite{Pearson:2018tkr}, the inner crust---including the pasta
phases---is obtained from a self-consistent Thomas--Fermi
calculation~\cite{Grill:2014aea}, and matter is homogeneous above
$n_B=0.067\fm^{-3}$. The table was taken from the \textsc{CompOSE}
repository~\footnote{\url{https://compose.obspm.fr/eos/217}}. DD2 reproduces
nuclear bulk properties at saturation ($E_0=-16.02\MeV$, $K=243\MeV$,
$J=31.7\MeV$, $L=55\MeV$) and yields a purely hadronic sequence with
$M_{\max}=2.42\,\Msun$ and $R_{1.4}=\revg{13.18}$~km, consistent with the
$2\,\Msun$ and NICER constraints; its tidal deformability, $\Lambda_{1.4}=690$,
lies somewhat above the GW170817 upper bound~\cite{Abbott:2018exr}.

\label{sec:quark}%
For the quark phase we adopt the three-flavor MIT bag model extended with a
repulsive vector
interaction~\cite{Farhi:1984qu,Franzon:2016urz,Lopes2021_I,Lopes2021_II},
defined by the Lagrangian density
\begin{align}
\mathcal{L} ={} &
  \sum_{q}\Bigl\{\bar{\psi}_{q}\bigl[i\gamma^{\mu}\partial_{\mu}
  - m_{q} + g\,\gamma^{\mu}V_{\mu}\bigr]\psi_{q} - B\Bigr\}\,
  \Theta\!\bigl(\bar{\psi}_{q}\psi_{q}\bigr)
\notag\\
& + \tfrac{1}{2}\,m_{V}^{2}\,V_{\mu}V^{\mu}
  + \sum_{l}\bar{\psi}_{l}\,\gamma_{\mu}
    \bigl(i\partial^{\mu}-m_{l}\bigr)\psi_{l}\,,
\label{eq:Lagrangian}
\end{align}
where the bag constant $B$ is the energy cost of the perturbative vacuum and
the vector-isoscalar field $V^\mu$ ($m_V=780\MeV$) couples universally to the
three flavors. In the mean-field approximation at $T=0$ the vector field
shifts the quark chemical potentials by
$\mu_q^{*}=\mu_q-G_V\,(n_u+n_d+n_s)$, with $G_V\equiv(g/m_V)^2$ (in units of
$\fm^{2}$), providing a purely repulsive contribution that stiffens the EOS
at high density; the resulting thermodynamics is standard and can be found in
Ref.~\cite{Lopes2021_II}. Matter is taken cold, neutrino-free,
$\beta$-equilibrated, and charge neutral, with current quark masses
$m_u=m_d=4\MeV$ and $m_s=93.4\MeV$. The model thus has two free parameters
with complementary roles: $B^{1/4}$ controls the deconfinement threshold and
$G_V$ the high-density stiffness, and their systematic variation is the basis
of the parameter-space exploration of Sec.~\ref{sec:param_space}.

\label{sec:hybrid}%
The two phases are joined by a Maxwell construction, appropriate in the limit
of large hadron--quark surface tension, in which the transition proceeds as a
discontinuous jump at a single coexistence pressure with no mixed  \cite{Lugones:2021tee}. The
coexistence point is fixed by the continuity of the baryon chemical potential
and of the pressure,
\begin{subequations}
\label{eq:Maxwell}
\begin{align}
\mu^{H}_B &= \mu^{Q}_B \equiv \mu_B^{\rm tr}\,,
\label{eq:Maxwell_mu}\\
p^{H}(\mu_B^{\rm tr}) &= p^{Q}(\mu_B^{\rm tr}) \equiv p_{\rm tr}\,,
\label{eq:Maxwell_p}
\end{align}
\end{subequations}
while the energy density jumps from $\varepsilon_-\equiv
\varepsilon^{H}(p_{\rm tr})$ to $\varepsilon_+\equiv
\varepsilon^{Q}(p_{\rm tr})$, with $\Delta\varepsilon=
\varepsilon_+-\varepsilon_->0$, and the baryon number density from
$n_B^{H}$ to $n_B^{Q}$, with $\Delta n_B \equiv n_B^{Q}-n_B^{H}>0$ (both
evaluated at $p_{\rm tr}$). The two jumps are not independent: since the
pressure and the chemical potential are continuous across the interface, the
thermodynamic identity $\varepsilon+p=\mu_B\,n_B$ applied to both phases at
coexistence yields the exact relation
\begin{equation}
\Delta\varepsilon = \mu_B^{\rm tr}\,\Delta n_B\,.
\label{eq:deps_dnB}
\end{equation}
Inside the star the transition manifests as a single spherical surface at
the radius $r_{\rm tr}$ where $p(r_{\rm tr})=p_{\rm tr}$, separating the
quark core ($r<r_{\rm tr}$) from the hadronic envelope ($r>r_{\rm tr}$). The
discontinuity $\Delta\varepsilon$ plays a central role in what follows: it
produces the kink in the mass--radius relation (Sec.~\ref{sec:MR}), enters
the junction condition for the tidal perturbation
[Eq.~\eqref{eq:junction_y}], and controls the dynamical stability of the
first hybrid configurations through the Seidov criterion
[Eq.~\eqref{eq:seidov}]. Throughout the paper the stiffness of each phase is
quantified by the squared speed of sound, $c_s^{2}\equiv dp/d\varepsilon$
(Sec.~\ref{sec:eos_cs}).

%--------------------------------------------------------------
\subsection{Stellar structure and notation}
\label{sec:stellar_structure}
%--------------------------------------------------------------
\label{sec:TOV}%
The macroscopic properties of each configuration are obtained by standard
methods~\cite{Haensel:2007yy}; here we fix the notation and collect only the
elements specific to hybrid stars. The
Tolman--Oppenheimer--Volkoff (TOV) equations~\cite{Oppenheimer:1939ne} are
integrated outward from the center with the EOS supplied in Maxwell form, so
that the density discontinuity at $r_{\rm tr}$ is handled automatically. The
integration yields the enclosed gravitational and baryon masses as running
functions of the radial coordinate,
\begin{align}
m_g(r) &= \int_0^{r} 4\pi r'^{2}\,\varepsilon(r')\,dr'\,,
\label{eq:Mgr}\\[2pt]
m_b(r) &= m_N\!\int_0^{r} 4\pi r'^{2}\,
\Bigl(1-\tfrac{2m_g(r')}{r'}\Bigr)^{\!-1/2} n_B(r')\,dr'\,,
\label{eq:MB}
\end{align}
where $\varepsilon(r)$ and $n_B(r)$ are the local energy and baryon number
densities and $m_N$ is the baryon mass unit, taken as the neutron mass,
$m_N=939\MeV$.

A hybrid star has two physically distinguished radii at which these running
functions are read off: the stellar surface $R$, where the pressure vanishes
($p=0$), and the hadron--quark interface $r_{\rm tr}$, where it equals the
transition value ($p=p_{\rm tr}$). Throughout the paper we evaluate every
macroscopic quantity at \emph{both} radii and distinguish the two evaluations
by an uppercase subscript: $S$ for the whole star ($r=R$) and $C$ for the
quark core ($r=r_{\rm tr}$). We further reserve a lowercase $m$ for the
running enclosed mass [Eqs.~\eqref{eq:Mgr}--\eqref{eq:MB}] and an uppercase
$M$ for the value it takes at a fixed radius. Combined with the lowercase
labels $g$ and $b$ for the gravitational and baryon masses, this gives
\begin{equation}
\begin{aligned}
M_{g,S} &\equiv m_g(R)\,, & \qquad M_{g,C} &\equiv m_g(r_{\rm tr})\,,\\
M_{b,S} &\equiv m_b(R)\,, & \qquad M_{b,C} &\equiv m_b(r_{\rm tr})\,,
\end{aligned}
\label{eq:masses}
\end{equation}
for the gravitational and baryon masses of the whole star and of its core.
Writing $R_S\equiv R$ and $R_C\equiv r_{\rm tr}$, the compactness at each
level is
\begin{equation}
\mathcal{C}_S\equiv\frac{M_{g,S}}{R_S}\,,\qquad
\mathcal{C}_C\equiv\frac{M_{g,C}}{R_C}\,.
\label{eq:core_quantities}
\end{equation}
The same $S/C$ labeling carries over to the tidal deformability and the
moment of inertia introduced next.

\label{sec:tidal}%
The dimensionless tidal deformability is
\begin{equation}
\Lambda_{S,C}  = \frac{2}{3}\,k_2^{S,C}\,\mathcal{C}_{S,C}^{-5}\,,
\label{eq:Lambda}
\end{equation}
where the tidal Love number $k_2^{S,C}$ is constructed from the metric
perturbation through the auxiliary function $y(r)\equiv rH'(r)/H(r)$, with
$H(r)$ the $\ell=2$ even-parity perturbation obeying the standard first-order
ODE integrated together with the TOV
system~\cite{Hinderer:2007mb,Postnikov:2010yn}. The two labels fix the radius
at which $y$ and the compactness are evaluated: the global value $k_2^{S}$
follows from $y(R)$ and $\mathcal{C}_S$, while the core value $k_2^{C}$
follows from the interior value $y_Q(r_{\rm tr})$, reached by the solution
within the quark core, and $\mathcal{C}_C$, the core boundary being treated
as an effective surface. At the sharp interface $y(r)$ is not continuous
but obeys the junction condition~\cite{Postnikov:2010yn,Takatsy:2020bnx}
\begin{equation}
y_H(r_{\rm tr}) = y_Q(r_{\rm tr})
  - \frac{4\pi\,r_{\rm tr}^{3}\,\Delta\varepsilon}
         {m_g(r_{\rm tr})+4\pi\,r_{\rm tr}^{3}\,p_{\rm tr}}\,,
\label{eq:junction_y}
\end{equation}
where $y_H(r_{\rm tr})$ is the value with which the integration proceeds
into the hadronic envelope. The second term on the right-hand side is
always negative, so that $y$ decreases on crossing the interface.
The same condition applies at the surface of a bare strange star,
where the density drops from $\varepsilon_s$ to zero, with
$\Delta\varepsilon=\varepsilon_s$ and $p=0$~\cite{Damour:2009vw,Postnikov:2010yn};
there it lowers $y(R)$ by $4\pi R^3\varepsilon_s/M$, close to $3$ at small
compactness, and raises $\Lambda$ by a large factor. We return to this
point in Sec.~\ref{sec:universal}, where it distinguishes the tidal
deformability of a quark core from that of a strange star.

Because the jump is applied where the integration leaves the quark
core, it modifies the solution in the hadronic envelope and hence
$\Lambda_S$, whereas the interior value $y_Q(r_{\rm tr})$ that enters
$\Lambda_C$ is unaffected. The direction of the effect follows from the
monotonic decrease of $k_2$ with $y(R)$ at fixed compactness: a lower
$y_H(r_{\rm tr})$ leads to a lower $y(R)$ and therefore to a larger
$\Lambda_S$ than would be obtained if the junction condition were
omitted~\cite{Han:2018mtj,Takatsy:2020bnx}. The magnitude of the effect
depends on the position of the interface. When the core occupies less than
about $40\%$ of the stellar radius, the perturbation of $y$ decays as the
integration proceeds through the envelope and $\Lambda_S$ changes by less
than $1\%$; for the most extended cores of our ensemble, with
$r_{\rm tr}\simeq0.9\,R$, omitting the jump would underestimate $\Lambda_S$
by $30$--$45\%$. This effect should not be confused with the reduction of
$\Lambda_S$ relative to a purely hadronic star of the same mass discussed
in Sec.~\ref{sec:universal}, which arises from the higher compactness and
central condensation of the hybrid configuration.

\label{sec:moment_of_inertia}%
The moment of inertia is computed in the slow-rotation
approximation~\cite{Hartle:1967he} from a third running function
$\mathcal{I}(r)$, which gives the moment of inertia interior to $r$ and solves
the first-order equation~\cite{Hu:2023vsq,Dong:2023vxv}
\begin{equation}
\frac{d\mathcal{I}}{dr} = \frac{8\pi}{3}\,r^4\,\varepsilon
  \!\left(1+\frac{p}{\varepsilon}\right)\!
  \left(1-\frac{5\mathcal{I}}{2r^3}
       +\frac{\mathcal{I}^2}{r^6}\right)\!
  \left(1-\frac{2m_g(r)}{r}\right)^{\!-1}\!,
\label{eq:dI_dr}
\end{equation}
integrated outward from $\mathcal{I}(0)=0$ alongside the TOV system. Unlike
the tidal function $y(r)$, whose equation involves the radial gradient of
$\varepsilon$ and therefore jumps at the transition, the source of
Eq.~\eqref{eq:dI_dr} depends on $\varepsilon$ and $p$ only algebraically; the
density discontinuity thus leaves $\mathcal{I}(r)$ continuous, and no junction
condition is required at $r_{\rm tr}$. Following the same $S/C$ convention we
read it off at the two radii, $I_S\equiv\mathcal{I}(R)$ for the whole star and
$I_C\equiv\mathcal{I}(r_{\rm tr})$ for the core, and form the dimensionless
moment of inertia
\begin{equation}
\bar I_{S,C} \equiv \frac{I_{S,C}}{\bigl(M_{g,S,C}\bigr)^{3}}\,.
\label{eq:Ibar}
\end{equation}
Equations~\eqref{eq:masses}--\eqref{eq:Ibar} thus supply, for every
configuration, a global ($S$) and a core ($C$) value of the mass, radius,
compactness, tidal deformability, and moment of inertia---the full set of
quantities with which universality is tested at both levels in
Sec.~\ref{sec:universal}.

%--------------------------------------------------------------
\subsection{Numerical method and validation}
\label{sec:numerics}
%--------------------------------------------------------------
The structure equations are integrated with an adaptive explicit
Runge--Kutta scheme (relative tolerance $10^{-9}$), with initial conditions
at $r_0=1$~m obtained from the regular series expansion of the TOV system,
$m(r_0)=\tfrac43\pi\varepsilon_c r_0^3$, $p(r_0)=p_c-\tfrac{2\pi}{3}
(\varepsilon_c+p_c)(\varepsilon_c+3p_c)r_0^2$, $y(r_0)=2$, and the
corresponding expressions for $m_b$ and $\mathcal I$. The hadron--quark
interface and the stellar surface are located as events of the integration,
i.e., as the roots of $p(r)=p_{\rm tr}$ and $p(r)=0$, rather than on a fixed
radial grid; the core quantities of Sec.~\ref{sec:stellar_structure} are read
off at the first event and the junction condition~\eqref{eq:junction_y} is
applied there before the integration proceeds through the envelope. Each
phase is described by its own tabulated EOS, and the speed of sound entering
the tidal equation is evaluated within each phase separately, so that the
density discontinuity enters only through the explicit junction
condition. 
Along each sequence the central pressure is sampled on
a grid with 1--10\% spacing, supplemented by 15 points in a geometric
progression between $p_{\rm tr}$ and the next grid point, so that cores as
small as a few metres are resolved. The core mass and radius satisfy
$M_{g,C}\simeq\tfrac43\pi R_C^3\,\varepsilon_+$ to better than $1\%$ for
cores below 100~m, as expected for a nearly uniform sphere. As a global
validation, the $\bar I$--$\Lambda$ relation of purely hadronic stars is
reproduced by our DD2 sequence to within $0.4\%$ of the fit of
Ref.~\cite{Yagi:2013bca}, the level at which that relation is known to hold
(Sec.~\ref{sec:universal}); for the canonical DD2 star we obtain
$R_{1.4}=13.18$~km and $\Lambda_{1.4}=690$.

%==============================================================
\section{Parameter space and viable hybrid configurations}
\label{sec:results}
%==============================================================

We begin by delimiting the region of the $(B^{1/4},G_V)$ plane that produces
physically acceptable hybrid stars, then summarize the resulting equations
of state and their speed of sound, and finally construct the corresponding
mass--radius sequences. The ensemble assembled here is the basis of the
universality analysis of Secs.~\ref{sec:universal}
and~\ref{sec:core_characterization}, and Sec.~\ref{sec:evanescent} classifies 
its members according to the size of the quark core they develop.

%==============================================================
\subsection{Exploration of the $(B^{1/4}, G_V)$ plane}
\label{sec:param_space}
%==============================================================

\begin{figure}[tb]
\includegraphics[width=0.99\columnwidth]{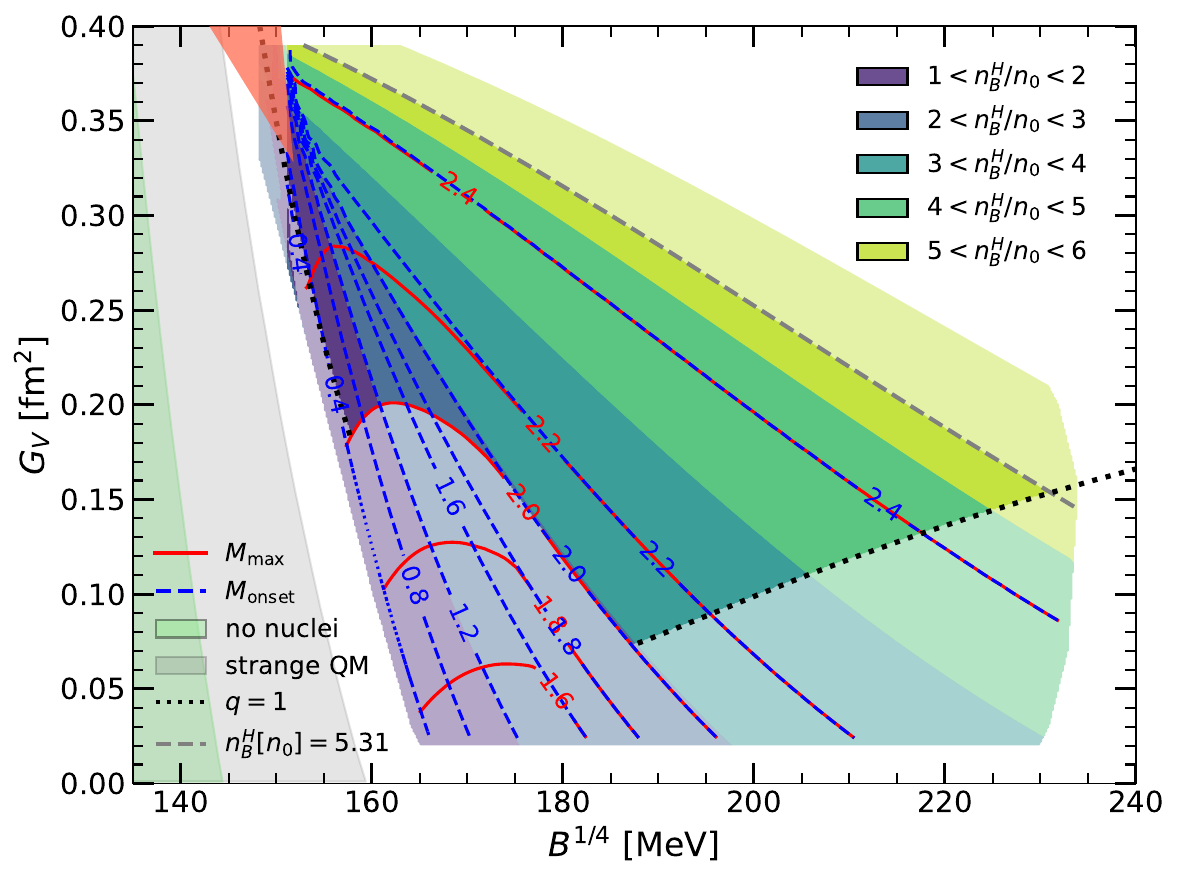}
\caption{Parameter space of the hybrid model in the $(B^{1/4},G_V)$ plane.
Red solid lines: contours of constant maximum gravitational mass $M_{\max}$
(in units of $M_\odot$). Blue dashed lines: contours of constant stellar
mass at the onset of the phase transition, $M_{\rm onset}\equiv M(p_{\rm tr})$
(in units of $M_\odot$). Colored bands: bins of the transition baryon
density on the hadronic side, $n_B^H/n_0$, where
$n_0=0.16\,\mathrm{fm}^{-3}$ is the nuclear saturation density. Gray dashed
line: $n_B^H=5.31\,n_0$, the central baryon density of the maximum-mass
configuration of the pure-DD2 sequence ($p_c=518\MeV\fm^{-3}$). 
Light-gray and light-green strips on
the left: regions in which three-flavor (``strange QM'') and two-flavor
(``no nuclei'') quark matter, respectively, would have an energy per baryon
below $930\,\MeV$. Coral shaded wedge: parametrizations with multiple
quark--hadron EOS crossings, also excluded from the analysis (see text).}
\label{fig:parameterspace}
\end{figure}

For each pair $(B^{1/4},\,G_V)$ we construct the hybrid EOS via the Maxwell
prescription of Sec.~\ref{sec:EOS}, solve the TOV equations, and extract the
maximum gravitational mass $M_{\max}$ and the corresponding radius
$R_{\max}$, the transition pressure $p_{\rm tr}$ and chemical potential
$\mu_B^{\rm tr}$, the baryon densities $n_B^H$ and $n_B^Q$ on either side of
the discontinuity with the associated jumps $\Delta n_B$ and
$\Delta\varepsilon$ [Eq.~\eqref{eq:deps_dnB}], and the stellar mass and radius
at the onset of the transition, $M_{\rm onset}\equiv M_{g,S}(p_{\rm tr})$ and
$R_{\rm onset}\equiv R(p_{\rm tr})$---the global mass and radius of the
configuration whose central pressure equals $p_{\rm tr}$. The scan
covers $G_V=0.01$--$0.40\fm^2$ in steps of $0.01\fm^2$ and
$B^{1/4}=145$--$234\MeV$ in steps of $1\MeV$, restricted to the pairs for
which the Maxwell construction yields a hadron--quark transition within
the hadronic table and above nuclear saturation density, $n_B^H>n_0$; this
leaves $2718$ candidate EOSs, which fill the non-rectangular region
summarized in Fig.~\ref{fig:parameterspace}.

%%%%%%%%%%%%%%%%%%%%%%%%%%%%%%%%%%%%%%% 
%   FIGURE 1
%%%%%%%%%%%%%%%%%%%%%%%%%%%%%%%%%%%%%%% 

Five conditions delimit the viable region:
\begin{itemize}

\item \emph{Maximum-mass constraint.} The contour $M_{\max}=2\,\Msun$
defines the lower boundary of the viable region, in compliance with the
heaviest precisely measured
pulsars~\cite{Demorest:2010bx,Antoniadis:2013pzd,Fonseca:2021wxt}.

\item \emph{Dynamical stability against the transition.} The Seidov
criterion~\cite{1971Seidov}
\begin{equation}
q \equiv \frac{\varepsilon_+/\varepsilon_-}
{\tfrac{3}{2}(1 + p_{\rm tr}/\varepsilon_-)} < 1
\label{eq:seidov}
  \end{equation}
guarantees that the configuration in which the quark core first appears is
gravitationally stable, so that the hybrid branch remains connected to the
hadronic one. The curve $q=1$ bounds the viable region along its lower-right
edge and, through a second segment where large $\Delta\varepsilon$ combines
with small $p_{\rm tr}$, along its upper-left edge as well.

\item \emph{Transition below the hadronic maximum.} Transitions with
$p_{\rm tr}>p_c^{\rm DD2}(M_{\max})=518\MeV\fm^{-3}$, i.e.\ $n_B^H>5.31\,n_0$, would occur within an
already dynamically unstable region of the hadronic branch: the maximum-mass
configuration is then purely hadronic and no stable hybrid configuration
exists. This condition closes the region from above and from the right.

\item \emph{Existence of a hadronic envelope.} Parametrizations for which
three-flavor quark matter is absolutely stable (energy per baryon below
$930\,\MeV$ at zero pressure, the Bodmer--Witten
regime~\cite{Bodmer:1971we,Witten:1984rs}) would produce self-bound strange
stars rather than hybrid stars, and those for which two-flavor quark matter
is absolutely stable would not allow ordinary nuclei in vacuum; both strips
are excluded from the analysis.

\item \emph{Multiple crossings.} In a wedge-shaped region of the plane (coral wedge
in Fig.~\ref{fig:parameterspace}) the quark and hadronic pressures run
nearly parallel as functions of the baryon chemical potential $\mu_B$ and
intersect at more than one point. The Maxwell conditions
[Eqs.~\eqref{eq:Maxwell}] then fail to single out a unique coexistence
pressure, leaving the hadron--quark transition ill-defined, so the
parametrizations falling in this region are excluded from the analysis.

\end{itemize}
Out of the $2718$ pairs scanned, $1112$ survive these constraints and
constitute the ensemble used throughout the remainder of this paper.

Inside the viable region the transition density spans
$n_B^H/n_0\sim1.5$--$5.3$, covering early- and late-onset configurations.
Both parameters delay the transition. At fixed $G_V$, a larger
$B^{1/4}$ raises the energy cost of the deconfined vacuum and shifts the
onset to higher density; at fixed $B^{1/4}$, a larger $G_V$ lowers the
quark pressure at a given baryon chemical potential, so that the quark
phase overtakes the hadronic one later and $n_B^H$ increases as well. The
bands of constant $n_B^H$ therefore run diagonally, from upper left to
lower right: a given transition density is reached with a lower
$B^{1/4}$ when $G_V$ is larger.

The blue dashed curves in Fig.~\ref{fig:parameterspace} are contours of constant $M_{\rm onset}$.
Since $M_{\rm onset}$ is the mass at which an infinitesimal quark core first
appears---the location of the kink in the mass--radius relation---its
separation from the $M_{\max}$ contour measures the mass interval available
for stable hybrid configurations: a large separation indicates quark cores
present well below $M_{\max}$, a small one a transition occurring close to
the maximum mass. 

A noteworthy feature of Fig.~\ref{fig:parameterspace} is that, in the lower-right portion of the diagram, the contours of constant $M_{\max}$ and constant $M_{\rm onset}$ approach each other asymptotically and align with the bands of constant $n_B^H$. This is no coincidence: as the quark cores become smaller and smaller, the star approaches a purely hadronic object whose central density is determined solely by the hadronic EOS, so it is natural that the convergence of the two families of contours takes place along curves of constant hadronic density. The fact that distinct $(B^{1/4},G_V)$ parametrizations sharing the same value of $M_{\max} \simeq M_{\rm onset}$ also share the same central density is a direct consequence of this: since the hadronic EOS is fixed, $M_{\rm onset}$ is in one-to-one correspondence with $p_{\rm tr}$ and hence with $n_B^H$, so a contour of fixed $M_{\rm onset}$ is automatically a contour of fixed $n_B^H$. In the limit of vanishing quark core the central density coincides with $n_B^H$, and the same property is inherited by $M_{\max}$ along the asymptotic portion of its contours.

%==============================================================
\subsection{Equation of state and speed of sound}
\label{sec:eos_cs}
%==============================================================

%%%%%%%%%%%%%%%%%%%%%%%%%%%%%%%%%%%%%%% 
%   FIGURE 2
%%%%%%%%%%%%%%%%%%%%%%%%%%%%%%%%%%%%%%% 
\begin{figure}[tb]
\centering
\includegraphics[width=0.90\columnwidth]{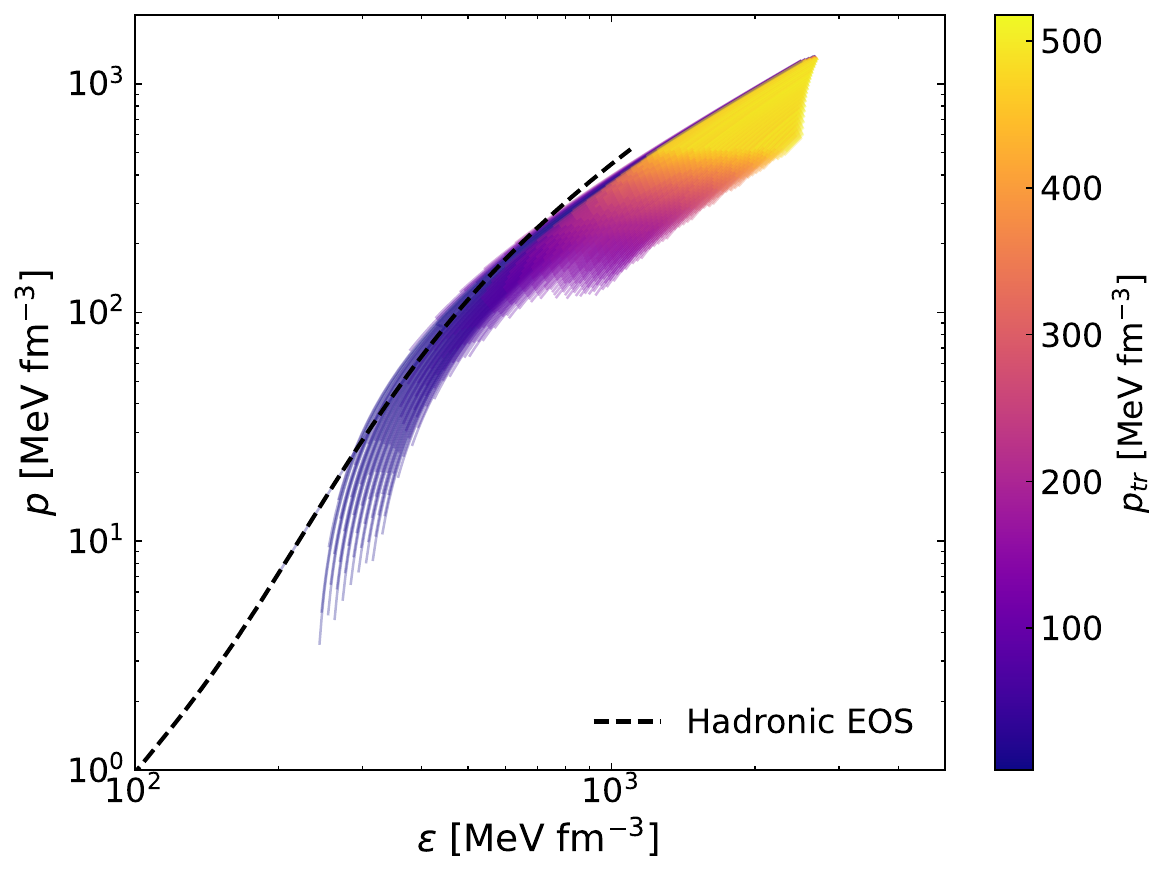}
\caption{Hybrid EOSs of the viable ensemble in the pressure--energy-density
plane, colored by the transition pressure $p_{\rm tr}$ of each EOS. Each
quark branch detaches from the hadronic DD2 EOS (dashed curve) at its
coexistence pressure and starts at
$\varepsilon_+=\varepsilon^{Q}(p_{\rm tr})$.}
\label{fig:cssquared}
\end{figure}

Figure~\ref{fig:cssquared} shows the equations of state of the viable
ensemble in the $p$--$\varepsilon$ plane. Each quark branch detaches from
the hadronic DD2 curve at its coexistence pressure, after the energy-density
jump $\Delta\varepsilon$, and runs with a nearly constant slope. The slope
is the squared speed of sound, $c_s^2=dp/d\varepsilon$, and the two phases
differ markedly in its behavior: across the ensemble the quark phase has
$c_s^{2}\simeq0.4$--$0.6$, increasing only mildly with density, whereas the
speed of sound of the hadronic phase rises steeply, approaching
$c_s^{2}\simeq0.75$ at the highest densities of the pure-DD2 sequence. For
the vast majority of the EOSs the quark branch therefore lies below the
hadronic curve at equal energy density, i.e.\ quark matter is softer than
DD2 matter, and the corresponding hybrid sequences of Sec.~\ref{sec:MR} are
more compact than the purely hadronic one. The exception is a small group
of EOSs with the lowest bag constants and the strongest vector couplings
($B^{1/4}\lesssim150\MeV$, $G_V\gtrsim0.3\fm^2$), in which the quark branch
starts at low density and, stiffened by the vector repulsion, overtakes the
DD2 curve over part of the density range; these are the few sequences of
Sec.~\ref{sec:MR} that extend to larger radii than the hadronic one.

The quark EOS is causal throughout: the vector repulsion keeps $c_s^{2}$
well above the conformal value but below unity at all densities. As in any
bag construction with constant $B$ and $G_V$, however, $c_s^2$ grows
monotonically with density and does not approach the conformal limit
$c_s^{2}\to1/3$ expected from the asymptotic freedom of
QCD~\cite{Kurkela:2009gj, Fraga:2013qra}. This has no bearing on the
present analysis, since the conformal regime sets in at baryon densities
$n_B\gtrsim40\,n_0$, far beyond the central densities of even the
maximum-mass configurations; the vector-MIT EOS is to be understood as an
effective description of the density window probed by hybrid stars.

%==============================================================
\subsection{Mass--radius sequences}
\label{sec:MR}
%==============================================================

%%%%%%%%%%%%%%%%%%%%%%%%%%%%%%%%%%%%%%% 
%   FIGURE 3
%%%%%%%%%%%%%%%%%%%%%%%%%%%%%%%%%%%%%%% 
\begin{figure}[tb]
\centering
\includegraphics[width=\columnwidth]{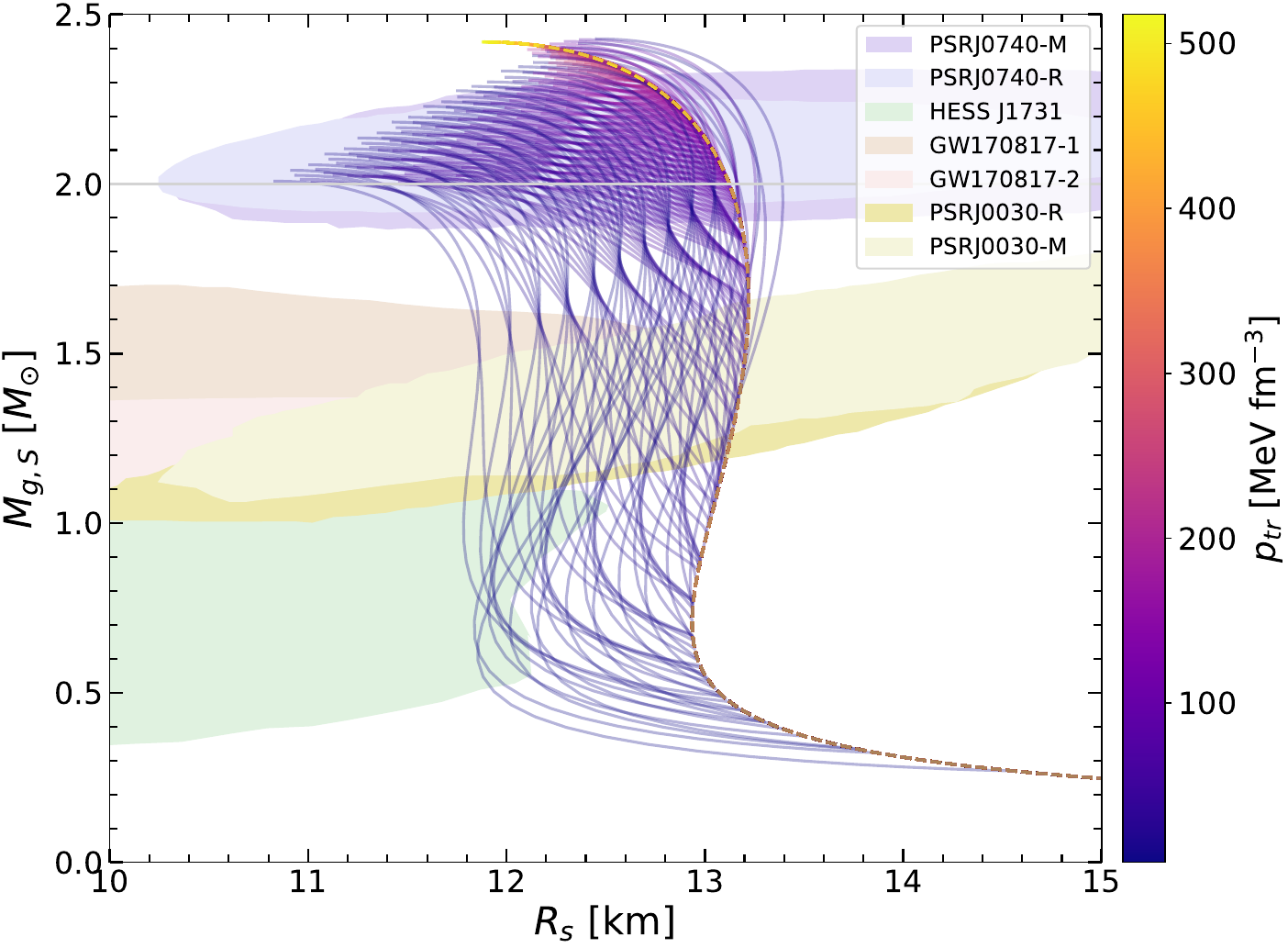}
\caption{Mass--radius sequences for the $1112$ viable hybrid EOSs, colored
by the transition pressure $p_{\rm tr}$ (in $\mathrm{MeV\,fm}^{-3}$). The
dashed line marks the purely hadronic DD2 sequence. Shaded regions show the
mass--radius constraints from
PSR~J0740$+$6620~\cite{Riley:2021pdl,Miller:2021qha},
PSR~J0030$+$0451~\cite{Riley:2019yda,Miller:2019cac}, the
tidal-deformability inference from
GW170817~\cite{TheLIGOScientific:2017qsa,Abbott:2018exr}, and
HESS~J1731$-$347~\cite{Doroshenko:2022nwp}.}
\label{fig:MvsR}
\end{figure}

The mass--radius sequences of the viable ensemble are displayed in
Fig.~\ref{fig:MvsR}, where the coloring by $p_{\rm tr}$ makes the role of
the transition location directly visible. The ensemble spans
$p_{\rm tr}\sim5$--$525\MeV\fm^{-3}$. EOSs with low $p_{\rm tr}$ develop a
quark core at low masses: the kink signaling the onset of the hybrid branch
appears in the lower part of the diagram, and the branch then extends over a
wide range of central densities, reaching radii as small as $\sim11$~km at
the maximum-mass end. EOSs with high $p_{\rm tr}$ develop a core only near
the maximum-mass configuration and track the purely hadronic DD2 curve over
almost the entire mass range. Intermediate values interpolate smoothly,
producing the characteristic fan: all sequences converge at low masses on
the common hadronic branch and spread at higher masses according to where
the transition takes place.

%%%%%%%%%%%%%%%%%%%%%%%%%%%%%%%%%%%%%%% 
% TABLE 1
%%%%%%%%%%%%%%%%%%%%%%%%%%%%%%%%%%%%%%% 
\begin{table*}[tb]
\centering
\caption{Representative members of the ensemble of 1112 viable hybrid EOSs, sampling the three
classes of core defined in Sec.~\ref{sec:evanescent}. For each EOS we list the two model parameters, the
quantities characterizing the first-order transition ($p_{\rm tr}$, $n_B^H/n_0$, $\Delta n_B$,
$\Delta\varepsilon$, the Seidov parameter $q$, and the interface ratio
$\xi=p_{\rm tr}/\varepsilon_+$ of Sec.~\ref{sec:smallcore}), the global properties of the stellar
sequence ($M_{\max}$, $R_{\max}$, $M_{\rm onset}$, $R_{\rm onset}$), the core mass and radius of
the maximum-mass configuration ($M_{g,C}^{\max}$, $R_C^{\max}$), and the compression ratio
$\eta^{\max}$ of Eq.~\eqref{eq:rho_eps}. The nuclear saturation density is
$n_0=0.16\,\mathrm{fm}^{-3}$. For EOSs with $M_{\max}-M_{\rm onset}<10^{-5}\,M_\odot$ the mass is
constant to within the numerical precision over a range of central pressures, so that the position
of the maximum along this plateau is not resolvable; for them the core mass and radius at the end
of the plateau are quoted as upper bounds.}
\label{tab:sets}
\footnotesize
\setlength{\tabcolsep}{3.2pt}
\begin{tabular}{c c c c c c c c c c c c c c c}
\hline\hline
$G_V$ & $B^{1/4}$ & $p_{\rm tr}$ & $n_B^H/n_0$ & $\Delta n_B$ & $\Delta\varepsilon$ & $q$ & $\xi$ & $M_{\max}$ & $R_{\max}$ & $M_{\rm onset}$ & $R_{\rm onset}$ & $M_{g,C}^{\max}$ & $R_C^{\max}$ & $\eta^{\max}$ \\
(fm$^{2}$) & (MeV) & (MeV\,fm$^{-3}$) &  & (fm$^{-3}$) & (MeV\,fm$^{-3}$) &  &  & ($M_\odot$) & (km) & ($M_\odot$) & (km) & ($M_\odot$) & (km) &  \\
\hline
0.10 & 190 & 154.7 & 3.30 & 0.34 & 463 & 0.950 & 0.150 & 2.151 & 12.98 & 2.151 & 12.98 & $<1.8\times10^{-4}$ & $<0.36$ & 1.000 \\
0.15 & 180 & 146.1 & 3.24 & 0.22 & 304 & 0.816 & 0.170 & 2.126 & 13.01 & 2.125 & 13.01 & $3.7\times10^{-3}$ & 1.05 & 0.989 \\
0.15 & 190 & 215.3 & 3.73 & 0.31 & 461 & 0.852 & 0.190 & 2.281 & 12.76 & 2.280 & 12.76 & $8.5\times10^{-5}$ & 0.27 & 0.999 \\
0.15 & 200 & 283.9 & 4.15 & 0.40 & 639 & 0.891 & 0.201 & 2.356 & 12.52 & 2.356 & 12.52 & $<1.9\times10^{-4}$ & $<0.33$ & 1.000 \\
0.15 & 210 & 353.6 & 4.53 & 0.49 & 837 & 0.929 & 0.207 & 2.394 & 12.30 & 2.394 & 12.30 & $<4.4\times10^{-5}$ & $<0.19$ & 1.000 \\
0.20 & 160 & 26.3 & 1.88 & 0.08 & 86 & 0.791 & 0.069 & 2.005 & 11.02 & 1.050 & 13.05 & 1.63 & 8.88 & 0.288 \\
0.20 & 170 & 124.6 & 3.06 & 0.13 & 172 & 0.716 & 0.180 & 2.076 & 12.47 & 2.045 & 13.09 & 0.37 & 5.04 & 0.744 \\
0.20 & 180 & 214.2 & 3.72 & 0.21 & 314 & 0.742 & 0.218 & 2.280 & 12.75 & 2.279 & 12.76 & $4.0\times10^{-3}$ & 1.03 & 0.987 \\
0.20 & 185 & 256.4 & 3.99 & 0.25 & 391 & 0.757 & 0.228 & 2.332 & 12.61 & 2.332 & 12.61 & $4.7\times10^{-4}$ & 0.48 & 0.997 \\
0.20 & 190 & 297.9 & 4.23 & 0.29 & 472 & 0.773 & 0.235 & 2.366 & 12.47 & 2.366 & 12.47 & $3.8\times10^{-4}$ & 0.43 & 0.997 \\
0.20 & 200 & 380.4 & 4.67 & 0.37 & 647 & 0.805 & 0.244 & 2.403 & 12.22 & 2.403 & 12.22 & $<7.8\times10^{-4}$ & $<0.51$ & 1.000 \\
0.25 & 160 & 69.2 & 2.52 & 0.05 & 62 & 0.656 & 0.147 & 2.124 & 11.65 & 1.691 & 13.22 & 1.22 & 7.83 & 0.402 \\
0.25 & 170 & 206.7 & 3.67 & 0.13 & 192 & 0.655 & 0.244 & 2.274 & 12.68 & 2.267 & 12.79 & 0.05 & 2.53 & 0.927 \\
0.25 & 175 & 261.7 & 4.02 & 0.17 & 264 & 0.668 & 0.261 & 2.338 & 12.57 & 2.337 & 12.59 & $6.7\times10^{-3}$ & 1.21 & 0.981 \\
0.25 & 180 & 313.3 & 4.32 & 0.21 & 339 & 0.682 & 0.271 & 2.376 & 12.42 & 2.375 & 12.42 & $1.7\times10^{-3}$ & 0.74 & 0.992 \\
0.25 & 190 & 412.4 & 4.83 & 0.28 & 498 & 0.708 & 0.284 & 2.410 & 12.14 & 2.410 & 12.14 & $7.4\times10^{-5}$ & 0.24 & 0.999 \\
0.30 & 160 & 185.7 & 3.53 & 0.06 & 87 & 0.585 & 0.262 & 2.284 & 12.32 & 2.228 & 12.87 & 0.37 & 4.96 & 0.717 \\
0.30 & 170 & 337.8 & 4.45 & 0.14 & 235 & 0.609 & 0.311 & 2.388 & 12.33 & 2.388 & 12.35 & $5.2\times10^{-3}$ & 1.09 & 0.982 \\
0.35 & 160 & 392.8 & 4.73 & 0.10 & 166 & 0.552 & 0.359 & 2.407 & 12.17 & 2.406 & 12.19 & $5.1\times10^{-3}$ & 1.07 & 0.981 \\
\hline\hline
\end{tabular}
\end{table*}

All $1112$ sequences satisfy $dM_{g,S}/d\varepsilon_c>0$ throughout the
hybrid branch, with $\varepsilon_c$ the central energy density of the
configuration. The Seidov criterion~\eqref{eq:seidov} guarantees this only at
the onset of the transition; we verified that the density jumps admitted by
the viable region are never large enough to destabilize the sequence further
along, so that no twin-star or disconnected branches appear and the model
populates the connected hybrid-star family exclusively.

The ensemble is broadly compatible with current multimessenger constraints:
it covers the regions inferred for PSR~J0740$+$6620 and is consistent with
PSR~J0030$+$0451; the largest-radius sequences lie close to, though not
always within, the GW170817 credibility region, as expected from the
statistical nature of that bound; and only the lowest-$p_{\rm tr}$,
smallest-radius sequences intersect the region inferred for
HESS~J1731$-$347, an inference that in any case relies on non-standard
atmosphere assumptions and should be regarded with caution. Canonical
stars retain the DD2 radius, $R_{1.4}=13.18$~km, for the $95\%$ of the
EOSs whose transition occurs above $1.4\,\Msun$, and are smaller, down to
$R_{1.4}=11.9$~km, for the earliest transitions; $M_{\max}>2\,\Msun$ holds
by construction.

Table~\ref{tab:sets} lists representative members of the ensemble,
sampling the extent of the viable region in both parameters and the three
classes of core defined in Sec.~\ref{sec:evanescent}, and makes
the systematic trends quantitative. For most of the EOSs listed the maximum and onset masses coincide to
three decimal places, a first indication of the plateau discussed in
Sec.~\ref{sec:evanescent}. At fixed $G_V$, increasing
$B^{1/4}$ pushes the transition to higher $\mu_B^{\rm tr}$ and
$p_{\rm tr}$, with growing jumps $\Delta n_B$ and $\Delta\varepsilon$;
both $M_{\max}$ and $M_{\rm onset}$ increase, and their difference
shrinks, the quantitative counterpart of the contour convergence
discussed in Sec.~\ref{sec:param_space}. At fixed $B^{1/4}$,
increasing $G_V$ also raises $p_{\rm tr}$ and $\mu_B^{\rm tr}$, while
stiffening the quark phase and raising $M_{\max}$. The jumps, however,
do not follow a common trend in absolute terms ($\Delta n_B$ decreases
only at the larger bag constants, and $\Delta\varepsilon$ shows no
systematic decrease); what decreases monotonically along this
direction is the \emph{relative} strength of the transition, as
quantified by the Seidov parameter $q$, and $M_{\rm onset}$
correspondingly approaches $M_{\max}$ here as well.

%==============================================================
\subsection{Evanescent and extended cores}
\label{sec:evanescent}
\label{sec:epsilon_ratio}
%==============================================================

%%%%%%%%%%%%%%%%%%%%%%%%%%%%%%%%%%%%%%% 
%   FIGURE 4
%%%%%%%%%%%%%%%%%%%%%%%%%%%%%%%%%%%%%%% 
\begin{figure}[tb]
\centering
\includegraphics[width=0.99\columnwidth]{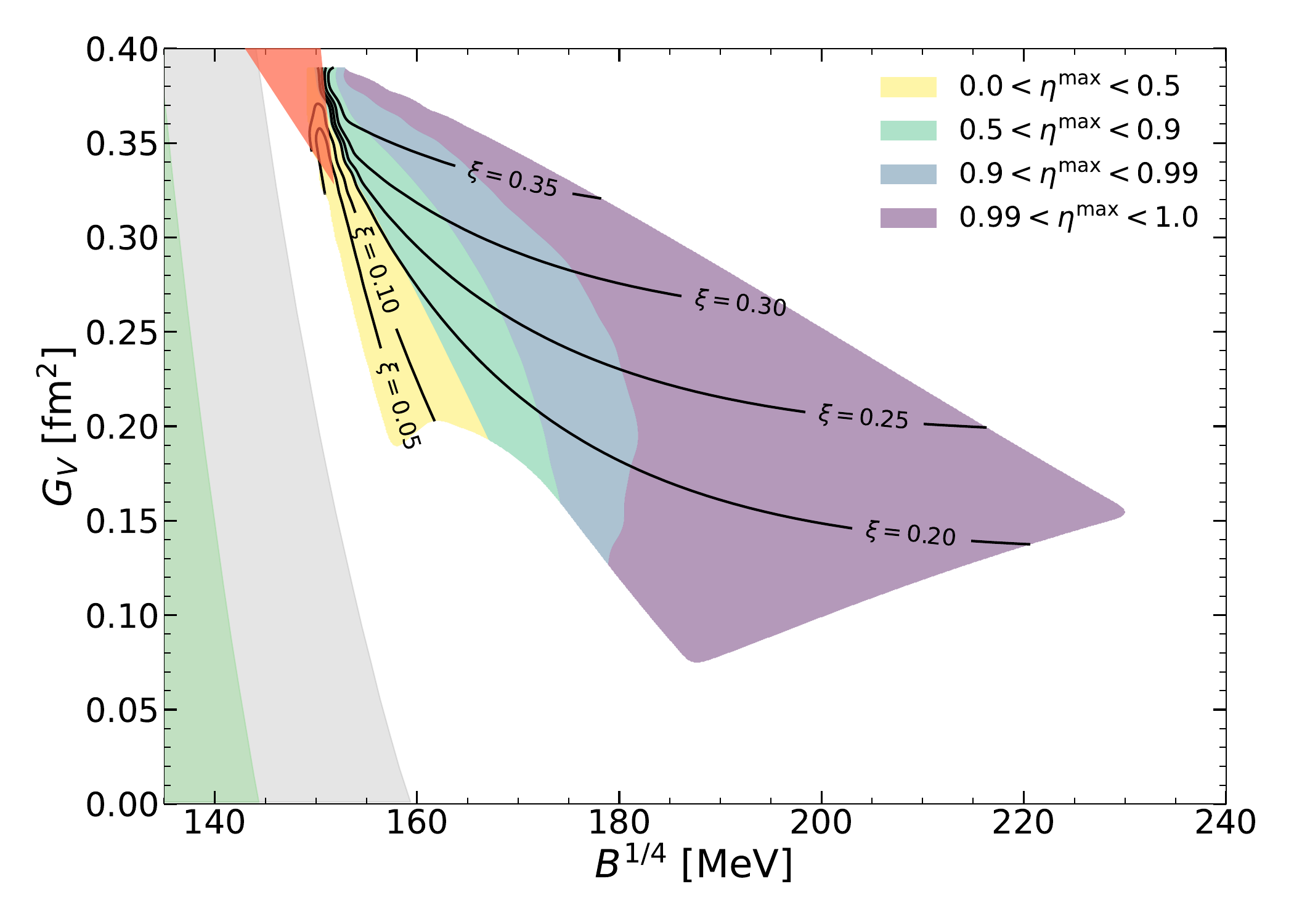}
\caption{Distribution of the energy-density compression ratio
$\eta^{\max}\equiv\varepsilon_+/\varepsilon_c^{\max}$
[Eq.~\eqref{eq:rho_eps}] over the viable region of the $(B^{1/4},G_V)$ plane,
for the three classes of core defined in the text: extended cores,
$\eta^{\max}<0.5$ (yellow); intermediate cores, $0.5<\eta^{\max}<0.9$
(green); and evanescent cores, $\eta^{\max}>0.9$, shown in two bins,
$0.9<\eta^{\max}<0.99$ (blue-gray) and $\eta^{\max}>0.99$ (purple), the
latter being the extremely evanescent cores, which appear only at the very
end of the stable sequence. Black curves: contours of constant
$\xi\equiv p_{\rm tr}/\varepsilon_+$, the interface ratio that organizes the
core relations of Sec.~\ref{sec:smallcore}. The light-gray and light-green
strips on the left are the ``strange QM'' and ``no nuclei'' exclusion
regions of Fig.~\ref{fig:parameterspace}.}
\label{fig:parameterspace_eps}
\end{figure}

For each viable EOS we define the energy-density compression ratio
\begin{equation}
\eta^{\max} \equiv
\frac{\varepsilon_+}{\varepsilon_c^{\max}}\,,
\label{eq:rho_eps}
\end{equation}
where $\varepsilon_+=\varepsilon^{Q}(p_{\rm tr})$ is the energy density of
the quark phase immediately after the transition
[Sec.~\ref{sec:EOS}] and $\varepsilon_c^{\max}$ is the central energy
density of the maximum-mass configuration of the sequence. By construction
$0<\eta^{\max}\leq 1$, and the ratio measures how much of the
energy-density range of the quark core actually develops beyond the
transition. The two limits are easy to interpret: when
$\eta^{\max}\to 1$ the transition pressure essentially coincides with
the central pressure of the maximum-mass star, so the quark core appears
only at the very end of the stable sequence and remains essentially
point-like (\emph{evanescent} cores); when $\eta^{\max}\ll 1$ the
transition occurs far below that pressure and the core develops over a wide
range of densities, occupying a substantial fraction of the star
(\emph{extended} cores). Because $\eta^{\max}$ is evaluated at the maximum-mass
configuration---which hosts the largest core along the sequence---it
characterizes a parametrization as a whole: a large $\eta^{\max}$ means
even the biggest core the sequence supports is tiny, while only EOSs with
small $\eta^{\max}$ develop sizable cores, near the high-mass end.

We group the viable EOSs into three classes according to
$\eta^{\max}$: extended cores, $\eta^{\max}<0.5$ ($108$ EOSs, $10\%$ of the
ensemble); intermediate cores, $0.5<\eta^{\max}<0.9$ ($114$ EOSs, $10\%$);
and evanescent cores, $\eta^{\max}>0.9$ ($890$ EOSs, $80\%$). Within the
last class we further single out the extremely evanescent cores,
$\eta^{\max}>0.99$ ($698$ EOSs, $63\%$ of the ensemble), whose quark core
appears only at the very end of the stable sequence.
Figure~\ref{fig:parameterspace_eps} shows the distribution of
$\eta^{\max}$ over the viable region with these bins. The dominant feature
is the extent of the evanescent class: for moderate-to-large bag constants
($B^{1/4}\gtrsim 175\MeV$ at any $G_V$, and the entire low-$G_V$ portion of
the region) the transition is late, the quark core appears close to
$M_{\max}$, and the energy-density range it samples is marginal. Extended cores with
$\eta^{\max}<0.5$ are confined to a narrow strip at the lowest viable
bag constants, $B^{1/4}\simeq 150$--$170\MeV$, and require at the same time a
sizable vector coupling, $G_V\gtrsim 0.2\fm^{2}$: a small $B$ brings the
deconfinement onset down to low pressures, while the strong vector repulsion
is what keeps the resulting long quark branch stiff enough to satisfy the
$2\,\Msun$ constraint. 
In Fig.~\ref{fig:MvsR}, the EOSs with low $p_{\rm tr}$ and
extended hybrid branches are exactly those with small $\eta^{\max}$,
while the sequences that track the DD2 curve until just below $M_{\max}$
populate the $\eta^{\max}>0.9$ region. 

Figure~\ref{fig:parameterspace_eps} also shows contours of constant
$\xi\equiv p_{\rm tr}/\varepsilon_+$, the ratio of the transition pressure
to the energy density of the quark phase at the interface, which will turn
out to be the parameter that controls the properties of the quark core
(Sec.~\ref{sec:smallcore}). It vanishes for a self-bound strange star,
whose quark surface lies at zero pressure, and it cannot exceed $\tfrac13$
in the bag model without vector interactions, for which
$p=(\varepsilon-4B)/3$; the vector repulsion stiffens the quark phase and
allows the larger values found at the high-$G_V$ end of the region. Across
the viable ensemble $\xi$ ranges from $0.014$ to $0.41$. Over most of the
plane the contours run nearly
horizontally, so that $\xi$ is set mainly by the vector coupling and grows
with it, while they crowd together in the extended-core strip, where
$p_{\rm tr}$ is small and $\xi$ falls rapidly toward zero: the EOSs with
the largest cores are also those with the smallest $\xi$.

The classification can be sharpened by looking at the extent of the hybrid
branch in mass. For 35\% of the viable EOSs the maximum mass exceeds the
onset mass by less than $10^{-5}\,M_\odot$, for 52\% by less than
$10^{-4}\,M_\odot$, and for 68\% by less than $10^{-3}\,M_\odot$: the hybrid
branch is a plateau on which the quark core grows while the mass stays
constant to within the numerical precision. The extremely evanescent
cores (purple region of Fig.~\ref{fig:parameterspace_eps}) are the EOSs in
this regime; their maximum-mass cores are smaller than
7\% of the stellar radius. Two consequences follow. First, for these EOSs the
transition occurs, for all practical purposes, at the maximum-mass
configuration, which is the strongest form of the masquerade: no global
observable of the stable sequence carries any imprint of the quark core.
Second, the position of the maximum along a plateau is not resolvable
numerically, so that the core of the maximum-mass configuration is an
ill-conditioned quantity for this class ($R_C\propto\sqrt{p_c-p_{\rm tr}}$
for small cores, so that the whole extent of the plateau maps onto a few
hundred metres in $R_C$); this is why Table~\ref{tab:sets} quotes only
upper bounds for $M_{g,C}^{\max}$ and $R_C^{\max}$ in such cases, and one
reason why the analysis of Sec.~\ref{sec:core_characterization} is carried
out on all stable configurations rather than on the maximum-mass ones
alone.

%==============================================================
\section{Universal and quasi-universal relations for star and core}
\label{sec:universal}
%==============================================================

%%%%%%%%%%%%%%%%%%%%%%%%%%%%%%%%%%%%%%% 
%   FIGURE 5
%%%%%%%%%%%%%%%%%%%%%%%%%%%%%%%%%%%%%%% 
\begin{figure*}[tb]
\centering
\includegraphics[width=0.90\textwidth]{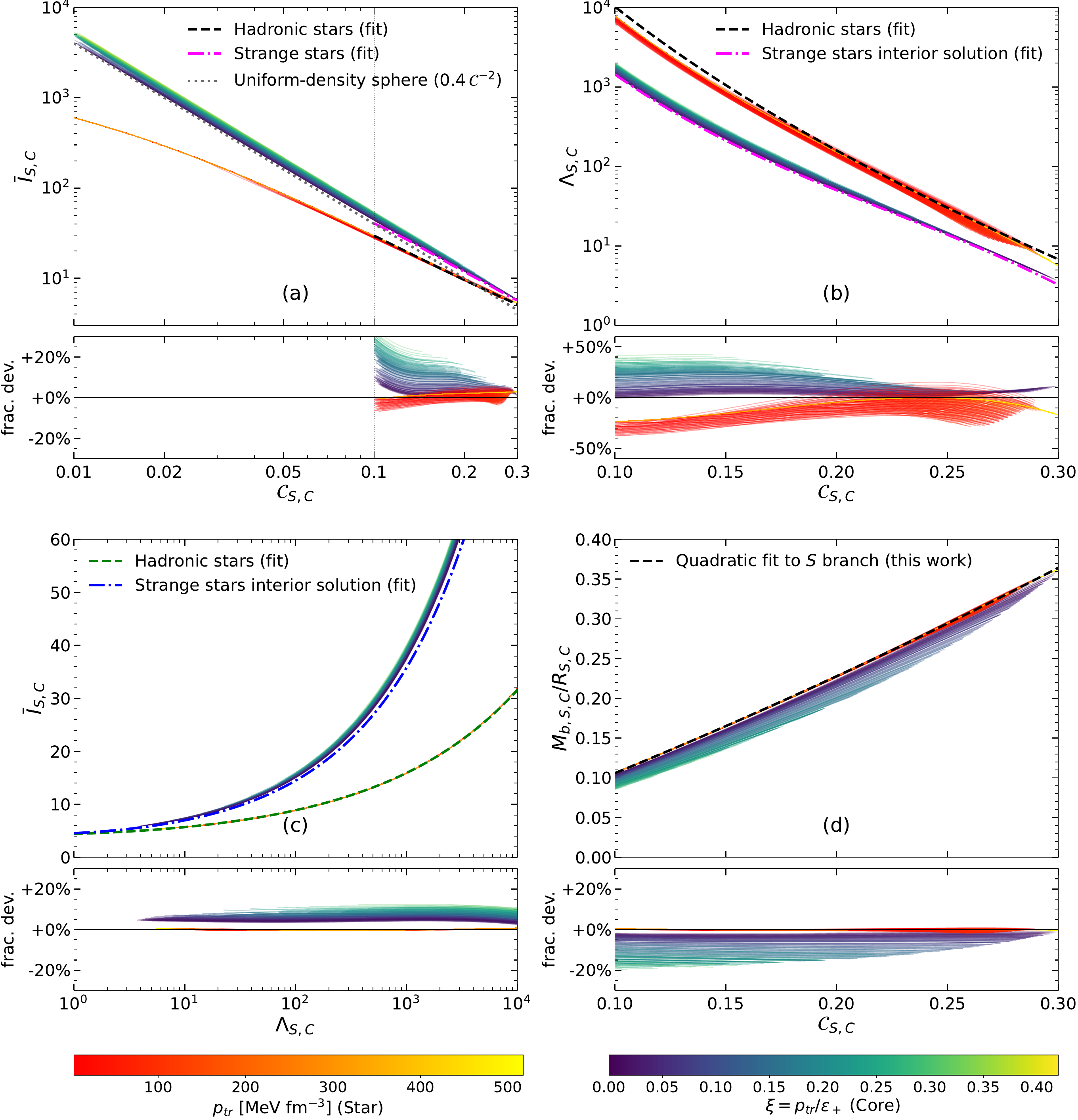}
\caption{Universal and quasi-universal relations for the $1112$ viable hybrid EOSs, with
every quantity evaluated at the stellar surface (label $S$) and at the
boundary of the quark core (label $C$): (a)~dimensionless moment of inertia
$\bar I$ [Eq.~\eqref{eq:Ibar}] versus compactness $\mathcal C_{S,C}$
[Eq.~\eqref{eq:core_quantities}], on
logarithmic axes; (b)~tidal deformability $\Lambda$ versus $\mathcal C$;
(c)~$\bar I$ versus $\Lambda$; (d)~baryon-mass compactness $M_b/R$ versus
$\mathcal C$. The $S$ branch is colored by the transition pressure
$p_{\rm tr}$ and the $C$ branch by the interface ratio
$\xi=p_{\rm tr}/\varepsilon_+$ [Eq.~\eqref{eq:sc_xi}]. Reference fits:
hadronic neutron stars~\cite{Breu:2016ufb,Maselli:2013mva,Yagi:2013bca}
(black and green dashed) and self-bound strange
stars~\cite{Lugones:2025lis,Yagi:2013bca} (magenta and blue dash-dotted),
drawn in their range of validity, the latter evaluated with the
interior solution for $y$ (see text); the Newtonian uniform-density sphere,
$\bar I=\tfrac25\mathcal C^{-2}$, in (a) (gray dotted); and a quadratic fit
to the $S$ branch in (d). The strips show the fractional deviation of every
configuration from the corresponding reference [the $S$ branch from the
hadronic fits, the $C$ branch from the strange-star fits; both from the
$S$-branch fit in (d)].   }
\label{fig:starcore}
\end{figure*}

A central question raised in the Introduction is whether the
quasi-universal relations that hold for ordinary neutron stars survive in
hybrid configurations, where the density
discontinuity could, in principle, spoil their EOS insensitivity---and,
beyond that, whether analogous relations also govern the quark core itself.
We address both questions by evaluating each correlation at two locations
of the same stellar model: at the surface,
where it probes the \emph{global} properties of the object, and at the core
boundary, where it probes the quark core in isolation.
We consider four correlations: $\bar I$--$\mathcal{C}$
[Eq.~\eqref{eq:Ibar}], $\Lambda$--$\mathcal{C}$, and
$\bar I$--$\Lambda$---a member of the original $I$--Love--$Q$
family~\cite{Yagi:2013bca,Yagi:2013awa,Yagi:2016bkt}---all known to be
(quasi-)universal for compact stars, together with the relation between
the baryon-mass compactness $M_b/R$ and the gravitational compactness
$\mathcal{C}$, whose degree of universality we assess here.
Each branch is compared against EOS-insensitive
reference fits from the literature: for $\bar I$--$\mathcal C$, the fit
calibrated on hadronic neutron stars by Breu and
Rezzolla~\cite{Breu:2016ufb}; for $\Lambda$--$\mathcal C$, the hadronic
relation of Maselli et al.~\cite{Maselli:2013mva}; for both, the fits
calibrated on self-bound strange quark stars in Ref.~\cite{Lugones:2025lis};
and for $\bar I$--$\Lambda$, the fits of Yagi and Yunes~\cite{Yagi:2013bca}.
The tidal fits of Ref.~\cite{Lugones:2025lis} were obtained from the
interior value of $y$ at the stellar surface, without the surface term of
Eq.~\eqref{eq:junction_y} that applies to a bare strange
star. They therefore describe a strange star
with the same inner boundary condition that defines the core quantities of
Sec.~\ref{sec:stellar_structure}, which is what makes them the appropriate
reference for the core branch.
The results are collected in Fig.~\ref{fig:starcore}, whose lower strips
show the fractional deviation of every configuration from the corresponding
reference.

%--------------------------------------------------------------
\subsection{Global observables}
\label{sec:global_obs}
%--------------------------------------------------------------

The global branches confirm, in compact form, what is known
from the literature. In Fig.~\ref{fig:starcore}(a) the $1112$ sequences pile onto the
hadronic $\bar I$--$\mathcal C$ fit over its whole range of validity,
$0.1\le\mathcal C_S\le0.3$, to within about $3\%$, the accuracy of the fit
itself, and with no trend in $p_{\rm tr}$. The reason is that the integrand
of Eq.~\eqref{eq:dI_dr} is proportional to $r^{4}\varepsilon$, so that
$I_S$ is dominated by the outer layers and inherits the scaling of a purely
hadronic configuration regardless of the presence of the core. 
The tidal deformability [Fig.~\ref{fig:starcore}(b)] is more sensitive to
the deep interior, yet most of the displacement of the $S$ branch from the
hadronic $\Lambda$--$\mathcal C$ fit of Maselli et al.~\cite{Maselli:2013mva}
is not produced by the quark core: it is already present in the purely
hadronic DD2 sequence, which lies about $20\%$ below that fit, and it is of
the size expected from the accuracy of the fit itself, which is quoted as
$2\%$ in $\mathcal C$ and, since $\Lambda$ falls roughly as
$\mathcal C^{-6}$, amounts to more than $10\%$ in $\Lambda$.
The imprint of the core is the
fan that opens at high compactness, where two effects of opposite sign are
at work: the dense core makes the hybrid star more centrally condensed than
a hadronic star of the same compactness, which lowers $k_2$, while the
junction condition [Eq.~\eqref{eq:junction_y}] raises $\Lambda_S$,
appreciably only for the most extended cores
(Sec.~\ref{sec:stellar_structure}). At fixed compactness the net displacement depends on
$\Delta\varepsilon$ and on the stiffness of the quark phase, not on
$p_{\rm tr}$ alone, which is why it is not resolved by the color coding. At
the level of individual $\bar I$--$\mathcal C$ or $\Lambda$--$\mathcal C$
measurements, then, a hybrid star of our ensemble is indistinguishable from
a purely hadronic one, in line with the few-percent agreement reported for
hybrid EOSs with strong first-order
transitions~\cite{Paschalidis:2017qmb}.

The $\bar I$--$\Lambda$ plane [Fig.~\ref{fig:starcore}(c)] shows the tightest correlation of
all. The $S$ branch follows the hadronic $I$--Love relation of
Ref.~\cite{Yagi:2013bca} to within $0.5\%$ over the entire ensemble, the
level at which that relation holds for purely hadronic stars, and it does so
also at the high-compactness end where the individual relations deviate: a
configuration whose $\Lambda_S$ lies below the hadronic $\Lambda(\mathcal C)$
curve has a $\bar I_S$ displaced by just the amount needed to keep
$(\Lambda_S,\bar I_S)$ on the hadronic curve. This mutual cancellation of
EOS-sensitive deviations is the hallmark of the $I$--Love--$Q$ relations,
and we find that it survives intact the presence of a density discontinuity
in the stellar interior, for the fluid cores considered here.

Figure~\ref{fig:starcore}(d) shows the baryonic compactness against the gravitational one. The
global branches form a narrow band, reproduced to within about $1\%$ by the
quadratic fit $M_{b,S}/R_S=0.721\,\mathcal C_S^2+1.004\,\mathcal C_S-0.002$,
so that this relation too is quasi-universal even when sizable quark cores
are present. The core branches fall systematically below it, by an amount
that grows with $\xi$; the reason is given in closed form below.

%--------------------------------------------------------------
\subsection{The quark core}
\label{sec:smallcore}
%--------------------------------------------------------------

Read at $r=r_{\rm tr}$, the same configurations abandon the hadronic
relations and organize along the strange-star fits of
Figs.~\ref{fig:starcore}(a)--\ref{fig:starcore}(c), with median
deviations of $6$--$9\%$ over the range of the fits, $0.1\le\mathcal C_C\le0.3$.
The deviations are largest at the low-compactness end of that range, where
they reach $30$--$40\%$ in Figs.~\ref{fig:starcore}(a) and
\ref{fig:starcore}(b), but remain below $12\%$ in Fig.~\ref{fig:starcore}(c).
{For the tidal deformability this agreement is with the strange star
read, like the core, from the interior solution; an actual strange star,
whose surface is a boundary with vacuum, would lie far above the core
branch in Figs.~\ref{fig:starcore}(b) and \ref{fig:starcore}(c)
(Sec.~\ref{sec:stellar_structure}). Three
features of the figure call for an explanation. First, the scatter is not
random but ordered by the interface ratio
\begin{equation}
\xi\equiv\frac{p_{\rm tr}}{\varepsilon_+}\,,
\label{eq:sc_xi}
\end{equation}
the cores with the smallest $\xi$ adhering to the fits and the deviation
growing with $\xi$. Second, below the range of the fits the core branch of
Fig.~\ref{fig:starcore}(a) follows the power law $\bar I_C\propto\mathcal C_C^{-2}$ of a
Newtonian sphere of uniform density (dotted line), displaced above it by an
amount that again grows with $\xi$. Third, the baryonic compactness of the
core [Fig.~\ref{fig:starcore}(d)] falls below the global relation by an amount that is, once
more, ordered by $\xi$. The three features have a common origin, which we
now make explicit by expanding the structure of the core in powers of its
compactness; the derivation is given in Appendix~\ref{app:smallcore}.

\paragraph*{Small-core expansion.}
A small core is one whose radius is short compared with the gravitational
length set by its own energy density, $r_{\rm tr}\ll\kappa^{-1/2}$ with
$\kappa\equiv\tfrac43\pi\varepsilon_+$. Since $m_g\simeq\kappa r^3$ inside
such a core, its compactness is $\mathcal C_C\simeq\kappa r_{\rm tr}^2$, and
$\mathcal C_C$ is the natural expansion parameter. A small core is far from
the high-density end of the quark EOS and probes it only in a neighborhood
of the transition point, so that its structure is fixed by
the quark-side interface quantities of Sec.~\ref{sec:EOS}---the pressure
$p_{\rm tr}$ and the energy density $\varepsilon_+$, which enter through
the ratio $\xi$ and the scale $\kappa$, and the chemical potential
$\mu_B^{\rm tr}$, which enters only through the baryon mass---together with
the squared sound speed of the quark phase at the interface,
$c_{s,+}^2\equiv(dp/d\varepsilon)_+$, which ranges from $0.38$ to $0.55$
across the ensemble. To zeroth order the core is a sphere
of uniform energy density $\varepsilon_+$ and uniform pressure
$p_{\rm tr}$. To first order the pressure and the density rise inward, and
the relativistic source terms of the TOV equation,
$\varepsilon+p=\varepsilon_+(1+\xi)$ and $m_g+4\pi r^3p=\kappa r^3(1+3\xi)$,
retain their dependence on $\xi$ no matter how small the core is, because
the pressure at its boundary is set by the envelope rather than by the core
itself. This is the essential difference between a quark core and a
self-bound strange star, for which $\xi=0$. Inserting the first-order
profiles into the structure equations of Sec.~\ref{sec:formalism} gives
\begin{subequations}
\label{eq:sc_all}
\begin{align}
M_C&=\kappa\,r_{\rm tr}^3\left[1+\frac{(1+\xi)(1+3\xi)}{5\,c_{s,+}^2}\,\mathcal C_C\right],
\label{eq:sc_M}\\[3pt]
\frac{M_{b,C}}{M_C}&=(1+\xi)\,\frac{m_N}{\mu_B^{\rm tr}}
\left[1+\left(\frac35-\frac{\xi(1+3\xi)}{5\,c_{s,+}^2}\right)\mathcal C_C\right],
\label{eq:sc_Mb}\\[3pt]
\bar I_C\,\mathcal C_C^{2}&=\frac25\,(1+\xi)\bigl[1+\beta\,\mathcal C_C\bigr],
\label{eq:sc_I}\\[3pt]
\Lambda_C\,\mathcal C_C^{4}&=\frac{2+\alpha}{15}\,,
\label{eq:sc_L}
\end{align}
\end{subequations}
with
\begin{equation}
\begin{aligned}
\alpha&=\frac17\left[1+33\xi+\frac{3(1+\xi)}{c_{s,+}^2}\right],\\
\beta&=\frac{6-2\xi}{7}-\frac{(1+3\xi)(2+7\xi)}{35\,c_{s,+}^2}\,,
\end{aligned}
\label{eq:sc_ab}
\end{equation}
where $M_C\equiv M_{g,C}$ and the omitted terms are of relative order
$\mathcal C_C^2$ in Eqs.~\eqref{eq:sc_M}--\eqref{eq:sc_I} and
$\mathcal C_C$ in Eq.~\eqref{eq:sc_L}. The core deformability is obtained,
as explained in Sec.~\ref{sec:stellar_structure}, from the interior value
$y_Q(r_{\rm tr})=2-\alpha\,\mathcal C_C$ of the tidal function
[Eq.~\eqref{eq:sc_y}].

%%%%%%%%%%%%%%%%%%%%%%%%%%%%%%%%%%%%%%% 
%   FIGURE 6: small core expansion
%%%%%%%%%%%%%%%%%%%%%%%%%%%%%%%%%%%%%%% 
\begin{figure*}[tb]
\centering
\includegraphics[width=0.93\textwidth]{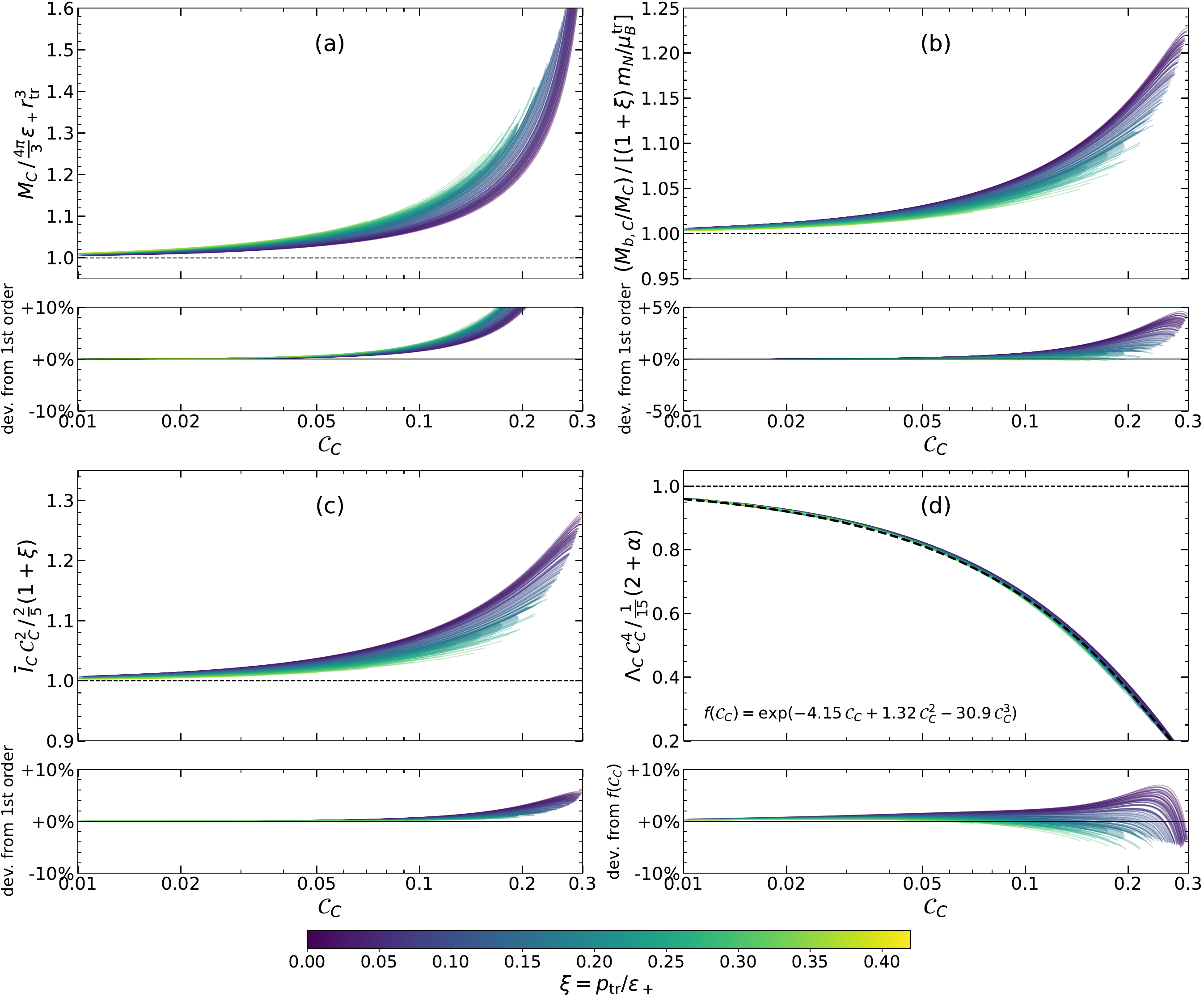}
\caption{Core quantities of all stable configurations of the $302$ EOSs
whose cores reach $\mathcal C_C>0.01$ along the stable branch, divided by
their zeroth-order small-core limits [Eqs.~\eqref{eq:sc_all}] and plotted
against the core compactness:
(a)~$M_C/\kappa r_{\rm tr}^3$; (b)~$(M_{b,C}/M_C)/[(1+\xi)m_N/\mu_B^{\rm tr}]$;
(c)~$\bar I_C\mathcal C_C^2/[\tfrac25(1+\xi)]$;
(d)~$\Lambda_C\mathcal C_C^4/[(2+\alpha)/15]$. Curves are colored by
$\xi=p_{\rm tr}/\varepsilon_+$. The black dashed curve in (d) is the
universal function $f(\mathcal C_C)$ of Eq.~\eqref{eq:sc_f}. The strips show the
fractional deviation from the first-order expressions
[Eqs.~\eqref{eq:sc_M}--\eqref{eq:sc_I}] in (a)--(c) and from
$f(\mathcal C_C)$ in (d).  }
\label{fig:smallcore}
\end{figure*}

\paragraph*{What the expansion says.}
Equations~\eqref{eq:sc_all} account for the three observations made above.
The leading term of $\bar I_C\mathcal C_C^2$ is the moment of inertia of a
uniform Newtonian sphere, $I=\tfrac25MR^2$, with the energy density
replaced by the inertial density $\varepsilon+p$: the offset of the core
branch above the dotted line of Fig.~\ref{fig:starcore}(a) is the factor
$(1+\xi)$, and it measures $\xi$ directly. The leading
term of the baryonic compactness,
$M_{b,C}/r_{\rm tr}=(1+\xi)(m_N/\mu_B^{\rm tr})\,\mathcal C_C$, is a
straight line whose slope is a property of the interface; since
$\mu_B^{\rm tr}$ exceeds $(1+\xi)\,m_N$ throughout the ensemble, the slope
is smaller than that of the star, increasingly so for larger $\xi$. The
baryonic--gravitational compactness relation is thus EOS-insensitive for
the whole star but not for the core in isolation, and its failure for the
core is not a breakdown of universality but a measurement of
$\mu_B^{\rm tr}$. Finally, with $\xi=0$ and $c_s^2=\tfrac13$, the interface
parameters of a bare strange star made of massless quarks,
Eqs.~\eqref{eq:sc_all} reduce to the small-compactness limit of the
strange-star relations evaluated, like those of the core, from the
interior solution: those relations are the $\xi\to0$
member of the family of analytic relations~\eqref{eq:sc_all}, and the
departures from them seen at low compactness in Fig.~\ref{fig:starcore}
measure $\xi$ itself. Two further remarks are in order.
The natural scaling of the core deformability is
$\Lambda_C\propto\mathcal C_C^{-4}$ rather than the $\mathcal C^{-5}$ of a
star with a finite Love number, because $k_2^C$ is of first order in the
compactness: it is evaluated from the interior solution,
$y_Q(r_{\rm tr})\to2$ as $\mathcal C_C\to0$, as appropriate for a core that
is not bounded by vacuum (Appendix~\ref{app:smallcore}). And the
first-order coefficient $\beta$ of $\bar I_C$ is small, between
$\simeq-0.1$ and $0.75$ across the ensemble, because its relativistic and
density-gradient contributions have opposite signs for all but the
smallest $\xi$, whereas the coefficient of $M_C$,
$(1+\xi)(1+3\xi)/(5c_{s,+}^2)$, grows with $\xi$ up to $\simeq1.7$; the
zeroth-order result for $\bar I_C$ therefore remains accurate to a few
percent up to $\mathcal C_C\simeq0.1$, while $M_C$ departs from its limit
earlier for most of the ensemble.

\paragraph*{Comparison with the ensemble.}
Figure~\ref{fig:smallcore}(a)--(d) shows the four core quantities of every stable
configuration of the $302$ EOSs whose cores reach $\mathcal C_C>0.01$,
divided by their zeroth-order limits, as functions of $\mathcal C_C$. All
curves tend to unity as $\mathcal C_C\to0$, at a rate that is linear in the
compactness and ordered by $\xi$, as the expansion prescribes, and the
first-order expressions reproduce the ensemble to better than $2.5\%$ up to
$\mathcal C_C=0.1$ (lower strips). The tidal deformability behaves
differently. Its first-order coefficient is large, so that the truncated
series is useful only below $\mathcal C_C\simeq0.05$; but once divided by
$(2+\alpha)/15$, the curves of all $302$ EOSs collapse onto a single
function of the compactness,
\begin{equation}
\begin{aligned}
\Lambda_C\,\mathcal C_C^{4}&=\frac{2+\alpha}{15}\,f(\mathcal C_C),\\
f(\mathcal C_C)&=\exp\!\bigl(-4.15\,\mathcal C_C+1.32\,\mathcal C_C^2-30.9\,\mathcal C_C^3\bigr),
\end{aligned}
\label{eq:sc_f}
\end{equation}
to within $2\%$ for $\mathcal C_C<0.1$ and a few percent over the whole
range up to $\mathcal C_C=0.3$. The entire EOS dependence of the core
deformability is thus carried by the single interface combination
$\alpha(\xi,c_{s,+}^2)$, while the dependence on the compactness is
universal.

\paragraph*{Discussion.}
Equations~\eqref{eq:sc_all}--\eqref{eq:sc_f} are not universal relations in
the usual sense: they contain the interface parameters explicitly. They are
rather the statement of what the core observables depend on, and of how
little. To first order, the mass, baryon mass, moment of
inertia, and tidal deformability of the core are fixed by its compactness
and by the state of matter on the quark side of the hadron--quark
interface---through $\xi$, $c_{s,+}^2$, and $\mu_B^{\rm tr}$---with no
separate dependence on $B$, $G_V$, or $\Delta\varepsilon$. This makes
precise the sense in which the core relations are ``strange-star-like'',
and it identifies what a measurement sensitive to the core alone would
deliver: the pressure at the interface relative to the energy density just
inside it, through $\xi$ and $\alpha$, rather than the model parameters
separately (Sec.~\ref{sec:inference}). The expansion can be carried to
second order, at the price of a fourth interface parameter,
$(dc_s^2/dp)_+$, and the universality of $f(\mathcal C_C)$ suggests that
the factorized structure persists well beyond the strict small-core regime;
we leave both developments, and the extension to other quark and hadronic
models, to future work.

%==============================================================
\section{The mass profile of hybrid stars and the limits of global-to-core inference}
\label{sec:core_characterization}
\label{sec:core_fraction}
\label{sec:inference}
%==============================================================

So far the quark core has been analyzed through dimensionless quantities
associated with its effective surface at $r=r_{\rm tr}$. We now turn to the
mass profile itself and ask how the mass enclosed by the phase boundary is
related to the radius of that boundary, for every stable configuration of
the ensemble. Neither quantity is observable, but their relation is a
stringent test of whether the normalized structure of hybrid stars is
universal, and it clarifies what the quark core does and does not add to
the structure of the star.

%%%%%%%%%%%%%%%%%%%%%%%%%%%%%%%%%%%%%%% 
%   FIGURE 7
%%%%%%%%%%%%%%%%%%%%%%%%%%%%%%%%%%%%%%% 
\begin{figure}[tb]
\centering
\includegraphics[width=0.99\columnwidth]{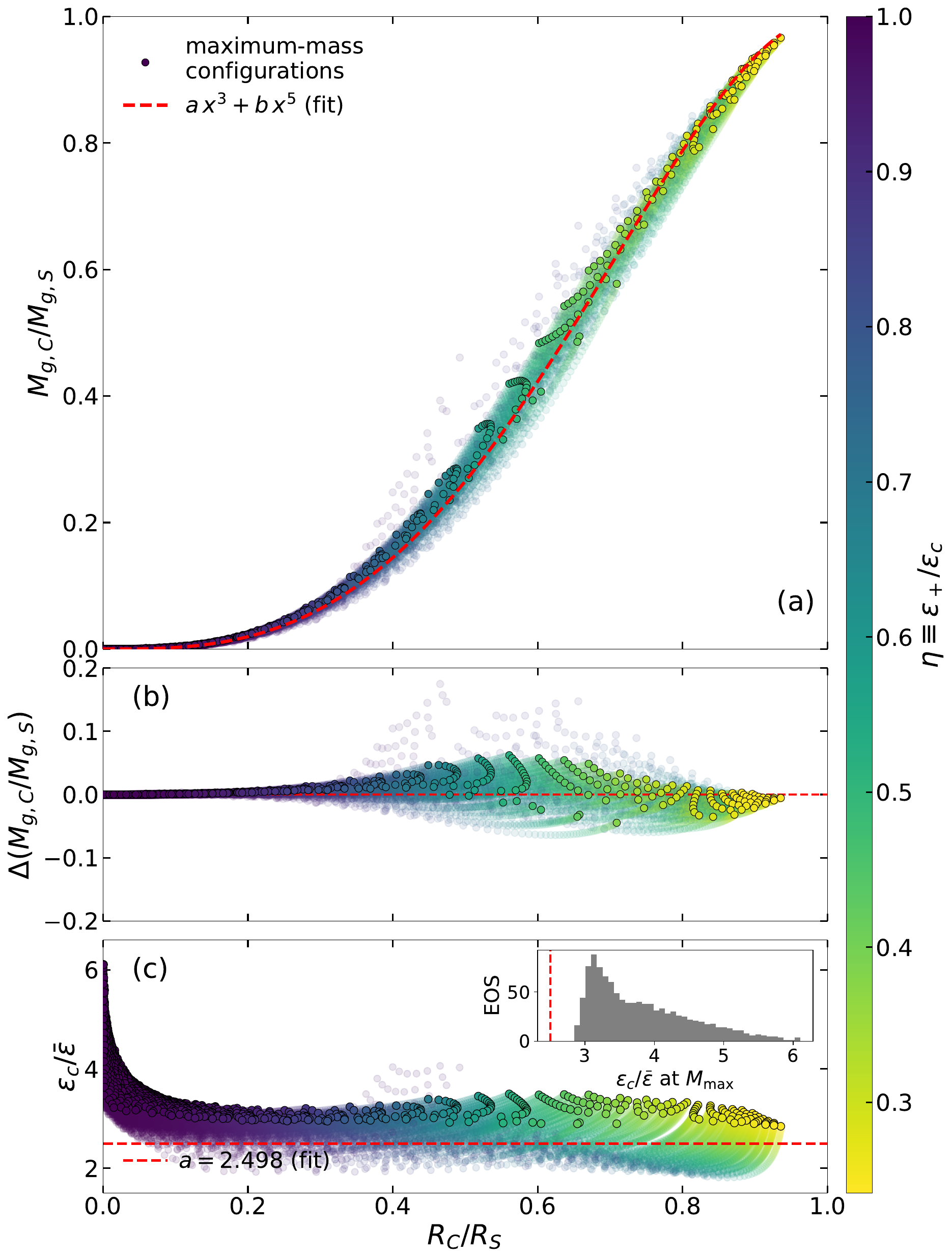}
\caption{(a) Fractional core mass $M_{g,C}/M_{g,S}$ versus fractional core
radius $R_C/R_S$ for the $54\,550$ stable configurations of the ensemble
(dots), with the maximum-mass configurations highlighted (circles with
black edge); the color gives the compression ratio
$\eta\equiv\varepsilon_+/\varepsilon_c$ of each configuration. The red
dashed curve is the fit of Eq.~\eqref{eq:core_fit}. (b)~Residuals
$\Delta\equiv M_{g,C}/M_{g,S}-f(R_C/R_S)$ with respect to the fit.
(c)~Central-to-mean energy-density ratio $\varepsilon_c/\bar\varepsilon$,
with $\bar\varepsilon\equiv3M_{g,S}/(4\pi R_S^3)$, for the same
configurations; the inset shows its distribution over the maximum-mass
configurations, and the dashed line marks the cubic coefficient $a$ of
Eq.~\eqref{eq:core_fit}.}
\label{fig:plot_ratios}
\end{figure}

For each viable EOS we sample 50 central pressures, equally spaced between
$p_{\rm tr}$ and the pressure of the maximum-mass configuration; 
for 21 EOSs with extremely evanescent cores the two pressures almost 
coincide and no configuration is sampled, which leaves
$54\,550$ stable hybrid configurations from 1091 EOSs. No additional
stability criterion is required, since $dM_{g,S}/d\varepsilon_c>0$ along
every hybrid branch (Sec.~\ref{sec:MR}). Figure~\ref{fig:plot_ratios}(a)
shows that the fractional core mass and the fractional core radius are
tightly correlated across the entire ensemble and collapse onto a single
curve, irrespective of the EOS and of the position of the configuration
along its sequence. The correlation is described by the two-parameter fit
\begin{equation}
\frac{M_{g,C}}{M_{g,S}} = f\!\left(\frac{R_C}{R_S}\right)\equiv
a \left(\frac{R_C}{R_S}\right)^{3}
+ b \left(\frac{R_C}{R_S}\right)^{5},
\label{eq:core_fit}
\end{equation}
with $a=2.498$ and $b=1-a=-1.498$, where the constraint $a+b=1$
ensures that a core filling the whole star encloses its whole mass. The
absolute residuals stay within $\simeq0.04$ in $M_{g,C}/M_{g,S}$, a bound
exceeded by only $1\%$ of the configurations
[Fig.~\ref{fig:plot_ratios}(b)]. The points span the full
range of core sizes, from point-like cores to the most extended one of the
ensemble, which reaches $R_C/R_S=0.94$ and $M_{g,C}/M_{g,S}=0.97$, and the
color coding by $\eta$ shows that the evanescent and the extended regimes,
and the full range of $(B^{1/4},G_V)$ that produces them, obey the same
relation.

The form of Eq.~\eqref{eq:core_fit} is suggested by the central expansion
of the mass profile: with $\varepsilon(r)=\varepsilon_c+\varepsilon_2r^2+O(r^4)$
near the center, $m(r)=\tfrac43\pi\varepsilon_c r^3+\tfrac45\pi\varepsilon_2
r^5+O(r^7)$, and normalizing $m$ by $M_{g,S}$ and $r$ by $R_S$ turns the
cubic coefficient into the central-to-mean density ratio
$\varepsilon_c/\bar\varepsilon$, with $\bar\varepsilon\equiv3M_{g,S}/(4\pi
R_S^3)$. Figure~\ref{fig:plot_ratios}(c) shows that  
$\varepsilon_c/\bar\varepsilon$ ranges from about
$3$ to $6$ among the maximum-mass configurations and is largest for the
evanescent cores, where $\varepsilon_c\simeq\varepsilon_+$ and the ratio
exceeds that of the underlying hadronic star by
$\Delta\varepsilon/\bar\varepsilon$. The fitted coefficients are
effective ones, determined by the intermediate and extended cores that
dominate the least-squares problem. Consistently, for small cores the
relation holds in absolute but not in relative terms: the
small-core limit of Eq.~\eqref{eq:sc_M} gives
$M_{g,C}/M_{g,S}=(\varepsilon_+/\bar\varepsilon)(R_C/R_S)^3$, with a
coefficient that is not universal but varies between about $2$ and $6$
across the ensemble [Fig.~\ref{fig:plot_ratios}(c)], while
$M_{g,C}/M_{g,S}$ itself stays below $0.02$. The universality of
Eq.~\eqref{eq:core_fit} is therefore a statement about the fractional core
mass to within a few percent of the stellar mass, not about the relative
size of tiny cores.

What Eq.~\eqref{eq:core_fit} describes deserves a comment. Since
$M_{g,C}$ is simply the mass enclosed by the sphere $r=r_{\rm tr}$, the pair
$(R_C/R_S,\,M_{g,C}/M_{g,S})$ is a point on the normalized mass profile
$m(r)/M_{g,S}$ versus $r/R_S$ of the star, and the collapse of
Fig.~\ref{fig:plot_ratios} states that this profile is nearly the same
function for all hybrid configurations of the ensemble, whatever the size of
the core; the quark core lies on the common curve because any sphere
interior to the star does. The profile of a hybrid star has a kink at
$r_{\rm tr}$, where the density jumps, but the enclosed mass is continuous
and the kink is invisible at the level of the scatter. The appropriate
reading is therefore not that quark cores possess a self-similar structure
of their own, but that the normalized mass profile of the star is
EOS-insensitive, across a first-order phase transition included, and that
the core-mass--core-radius relation is the form this insensitivity takes
when the profile is read at the phase boundary. Unlike the
core relations of Sec.~\ref{sec:smallcore}, which carry the interface
parameters explicitly, it involves no microphysical input at all, and it
holds throughout the stable hybrid branches, from the onset of the
transition to the maximum mass.

The relations found here connect core quantities to each other and
global quantities to each other. Whether a relation of a third kind exists,
with a core quantity on one side and a quantity measurable at the surface
on the other, is the possibility left open in the Introduction.
We searched systematically among the usual macroscopic
quantities and their combinations, but found no star--core relation whose
tightness is comparable to that of the relations obeyed by the core and by
the star separately. The information about the core carried by global
observables resides in their deviations from the purely hadronic star of
the same mass, which are below $0.6\%$ in radius and $6\%$ in $\Lambda_S$
for cores smaller than one fifth of the stellar radius and which, for
larger cores, depend on $\Delta\varepsilon$ and on the stiffness of the
quark phase as much as on the core size. Within this class of models,
static global observables therefore cannot be inverted to yield the core,
and Eq.~\eqref{eq:core_fit} is a constraint that any future inference must
satisfy rather than an inference tool. Breaking the degeneracy requires an
observable that responds directly to the interface, most naturally its
$g$-mode (Sec.~\ref{sec:conclusions}), and the small-core relations of
Sec.~\ref{sec:smallcore} specify what such an observable would measure: the
interface ratio $\xi$ and the combination $\alpha(\xi,c_{s,+}^2)$, rather
than the bag constant or the vector coupling separately.

%==============================================================
\section{Conclusions}
\label{sec:conclusions}
%==============================================================

We have performed a systematic study of hybrid stars whose quark cores are
described by the three-flavor MIT bag model with repulsive vector
interactions, matched to the hadronic DD2 EOS through a Maxwell
construction. Scanning the $(B^{1/4},G_V)$ plane and imposing dynamical
stability at the onset of the transition, a maximum mass above $2\,M_\odot$,
the existence of a hadronic envelope, a transition below the maximum of the
hadronic sequence, and a well-defined Maxwell construction, we obtained an
ensemble of 1112 viable hybrid EOSs whose transition densities span
$n_B^H\simeq1.5$--$5.3\,n_0$. For every stable configuration we evaluated
the mass, radius, compactness, moment of inertia, and tidal deformability
twice---at the stellar surface and at the hadron--quark interface---with a
numerical scheme that resolves cores as small as a few metres, turning each
stellar model into a simultaneous probe of global and core-level
(quasi)-universality. Five results emerge:

\begin{itemize}

\item Read at the surface, the sequences follow the universal relations
of purely hadronic neutron stars: $\bar I_S$ agrees with the hadronic
$\bar I$--$\mathcal C$ fit to within the accuracy of the fit itself (about
$3\%$), the $I$--Love relation holds to $0.5\%$ over the entire ensemble,
and the departures of $\Lambda_S$ from the hadronic
$\Lambda$--$\mathcal C$ reference are dominated by the offset of the DD2
sequence itself with respect to that reference, not by the quark core.
EOS-insensitive relations thus survive the density discontinuity, and at
the level of these relations a hybrid star of the ensemble is
indistinguishable from a hadronic one.

\item Read at the core boundary, the same configurations depart from
the hadronic relations and follow the scaling laws of self-bound strange
stars with median deviations of $6$--$9\%$, with deviations
ordered by the interface ratio $\xi=p_{\rm tr}/\varepsilon_+$ and
vanishing as $\xi\to0$, even though the core is held together by the
weight of the envelope rather than by self-binding. For the tidal
deformability, the comparison is made with the tidal function of both
objects evaluated just inside the boundary, without the surface term that
a density drop to vacuum would add. We have traced this behavior to its
origin by expanding the structure of the core in powers of its compactness. 
To first order, the mass, baryon mass, moment of inertia, and tidal
deformability of the core are given by closed-form expressions
[Eqs.~\eqref{eq:sc_all}] in which the quark EOS enters only through the
state of matter on the quark side of the interface---$\xi$, $c_{s,+}^2$,
and $\mu_B^{\rm tr}$---with no separate dependence on $B$, $G_V$, or
$\Delta\varepsilon$. These expressions reproduce the ensemble to within a
few percent for core compactness below $0.1$, and the tidal deformability
of the core factorizes into an interface amplitude, $(2+\alpha)/15$, and a
universal function of the compactness that holds to a few percent over the
whole ensemble. The strange-star relations are the $\xi\to0$ member of
this family, so that both the resemblance between quark cores and strange
stars and its degradation with the transition pressure are quantitative
consequences of the pressure at the interface.

\item The fractional core mass and the fractional core radius of all
stable hybrid configurations of the ensemble ($54\,550$ stellar models,
sampled along the hybrid branch of each EOS as described in
Sec.~\ref{sec:core_characterization}) fall on a single cubic-plus-quintic
curve [Eq.~\eqref{eq:core_fit}], to within $\simeq0.04$ of the
stellar mass (a bound only rarely exceeded). This curve is the normalized
mass profile $m(r)/M$ of the star read at the phase boundary, and its
universality states that the profile is EOS-insensitive across the phase
transition included. The cubic coefficient of the fit is an effective one
and should not be identified with the central-to-mean density ratio, which
varies between $3$ and $6$ across the ensemble and is largest for
evanescent cores; the small-core limit of the relation,
$M_{g,C}/M_{g,S}\to(\varepsilon_+/\bar\varepsilon)(R_C/R_S)^3$, makes this
explicit.

\item Hybrid stars that masquerade as hadronic ones~\cite{Alford:2004pf}
are, in this class of models, the default rather than a cautionary special
case. Evanescent cores dominate the viable $(B^{1/4},G_V)$ plane: for about
half of the EOSs the hybrid branch spans less than $10^{-4}\,M_\odot$ in
mass, so that the transition occurs, for all practical purposes, at the
maximum-mass configuration. Sizable cores require a bag constant low enough
to bring the deconfinement onset down to low pressures and a vector
repulsion strong enough to keep the resulting long quark branch above
$2\,M_\odot$, and are confined to a narrow strip at
$B^{1/4}\simeq150$--$170\MeV$ and $G_V\gtrsim0.2\fm^2$. This is in line with
the observation of Ref.~\cite{Komoltsev:2024fop} that weak transitions are
precisely the ones no current astrophysical datum can disfavor. We stress
that ``default'' refers to the extent of the corresponding region of the
parameter plane, not to a probability: there is no physical measure on the
model parameters, and the dominance of evanescent cores implies nothing
about the likelihood that nature realizes them.

\item The possibility of reading the core from the surface receives a
negative answer. A systematic search among the usual
global observables and their combinations finds no star--core relation of
a tightness comparable to the relations obeyed by the star and by the core
separately: the information about the core carried by global observables
resides in their deviation from the purely hadronic star of the same mass,
which is small for small cores and degenerate with the energy-density jump
and the stiffness of the quark phase for large ones. The universal
relations found here are thus relations among core quantities and among
global quantities, not between the two. Breaking the degeneracy requires an
observable that responds directly to the interface, such as the $g$-mode it
sources, and the small-core relations specify what it would measure: the
pressure at the interface relative to the energy density just inside it.

\end{itemize}

These conclusions are robust against the model choices for different
reasons at the two levels. The global relations are compared with
EOS-insensitive fits calibrated on broad samples of
EOSs~\cite{Yagi:2013bca,Breu:2016ufb,Maselli:2013mva}, so that replacing DD2
by another viable hadronic model would move our sequences within the band
traced by those samples, not away from it; the only global feature that
depends on the strength of the transition, the displacement of $\Lambda_S$
produced by the dense core and by the junction condition, is one that our
scan varies systematically from nearly vanishing jumps to the largest ones
allowed by stability. The core relations rest on the small-core expansion,
which assumes only a sharp interface and a regular quark EOS and is
therefore independent of the vector-MIT model, although it does not apply
without modification to a Gibbs mixed phase, where neither $\xi$ nor
$c_{s,+}^2$ is defined. The universality of the mass
profile follows from structural properties---a boundary at finite energy
density enclosing an interior of limited density contrast---that are
realized across the whole scan and are not tied to the specifics of the
model; alternative quark models producing cores with the same structure
should populate the same bands, consistently with the quasi-universality
reported for hybrid configurations built from different
microphysics~\cite{Paschalidis:2017qmb,Roy:2025rdy}.

The robustness is structural rather than quantitative, however. In the
Maxwell construction the hadronic side alone fixes the relation between
$p_{\rm tr}$, $n_B^H$, and $\mu_B^{\rm tr}$, so that the numerical boundaries
of the viable region, and in particular the fraction of parameter space
occupied by evanescent cores, inherit the choice of DD2; a softer hadronic
EOS would advance the onset and reduce that fraction. Extending the scan to
a representative set of hadronic EOSs is thus the natural next step,
together with confronting the framework with the interface $g$-mode
spectrum.

\section*{Acknowledgements}
This work has been supported in part by Consejo Nacional de Investigaciones Cientificas y
Tecnicas and Agencia Nacional de Promocion Cientifica y Tecnologica (Argentina), under Grants
No. PIP2022-GI-11220210100150CO and by the National University of La Plata (Argentina), Project No. X960.
G.~L. acknowledges partial support from Conselho Nacional de Desenvolvimento Cient\'{\i}fico e
Tecnol\'ogico (CNPq, Brazil) under Grant No.~313624/2026-7.
%==============================================================

%==============================================================
\appendix
\section{Derivation of the small-core expansion}
\label{app:smallcore}
%==============================================================

This appendix derives the first-order expressions of
Sec.~\ref{sec:smallcore}. The idea is to expand the structure equations of
Sec.~\ref{sec:formalism} for a core so small that its self-gravity only
slightly perturbs a sphere of uniform density and pressure, and to keep only
the terms linear in its compactness. The quark EOS enters
only through the quantities evaluated on the quark side of the interface
introduced in Sec.~\ref{sec:EOS}---the transition pressure $p_{\rm tr}$, the
energy density $\varepsilon_+$, the baryon density $n_B^Q$, and the chemical
potential $\mu_B^{\rm tr}$---together with the squared sound speed of the
quark phase at the interface, $c_{s,+}^2\equiv(dp/d\varepsilon)_+$. The
first two appear in the combinations
\begin{equation}
\kappa\equiv\frac{4\pi}{3}\,\varepsilon_+,\qquad
\xi=\frac{p_{\rm tr}}{\varepsilon_+}\quad[\text{Eq.~\eqref{eq:sc_xi}}],
\label{eq:app_def}
\end{equation}
so that $4\pi\varepsilon_+=3\kappa$, an identity used repeatedly below.
As in Sec.~\ref{sec:smallcore}, $M_C\equiv M_{g,C}$ denotes
the gravitational mass of the core.

\subsection{Expansion parameter and radial averages}

For a core of radius $r_{\rm tr}\ll\kappa^{-1/2}$ the energy density
is, to zeroth order, the uniform value $\varepsilon_+$, and the enclosed mass
and the local compactness are
\begin{equation}
m_g(r)\simeq\kappa r^3,\qquad
\frac{m_g}{r}\simeq\kappa r^2\le\kappa r_{\rm tr}^2\simeq\mathcal C_C .
\label{eq:app_m0}
\end{equation}
Every factor $\kappa r^2$ therefore counts as one order in $\mathcal C_C$:
the zeroth-order core has uniform $\varepsilon_+$ and uniform pressure
$p_{\rm tr}$, and the first-order result keeps the terms linear in
$\kappa r^2$. Since the exact compactness $\mathcal C_C=M_C/r_{\rm tr}$
differs from $\kappa r_{\rm tr}^2$ by a relative $O(\mathcal C_C)$
[Eq.~\eqref{eq:sc_M}], the two are interchangeable in first-order terms;
the results below are written with $\kappa r_{\rm tr}^2$ and become
Eqs.~\eqref{eq:sc_all} upon replacing it by $\mathcal C_C$. The
radial integrals that appear reduce to the averages
\begin{equation}
\begin{aligned}
\langle r^2\rangle_n&\equiv\frac{\int_0^{r_{\rm tr}}r^{n+2}dr}{\int_0^{r_{\rm tr}}r^{n}dr}
=\frac{n+1}{n+3}\,r_{\rm tr}^2,\\
\langle r_{\rm tr}^2-r^2\rangle_n&=\frac{2}{n+3}\,r_{\rm tr}^2,
\end{aligned}
\label{eq:app_avg}
\end{equation}
with $n=2$ for the masses and $n=4$ for the moment of inertia.

\subsection{Pressure and density profiles}

The pressure profile follows from the TOV equation,
\begin{equation}
\frac{dp}{dr}=-\frac{(\varepsilon+p)\,(m_g+4\pi r^3p)}{r^2\,(1-2m_g/r)}\,.
\label{eq:app_TOV}
\end{equation}
Its right-hand side is already of first order, so its factors can be
evaluated with the zeroth-order profiles,
\begin{equation}
\begin{gathered}
\varepsilon+p\to\varepsilon_+(1+\xi),\qquad
m_g+4\pi r^3p\to\kappa r^3(1+3\xi),\\
\Bigl(1-\frac{2m_g}{r}\Bigr)^{-1}\to1,
\end{gathered}
\label{eq:app_src}
\end{equation}
which turns Eq.~\eqref{eq:app_TOV} into
$dp/dr=-\kappa\varepsilon_+(1+\xi)(1+3\xi)\,r$. Integrating from the
interface, where $p=p_{\rm tr}$, and linearizing the EOS about it,
\begin{align}
p(r)&=p_{\rm tr}+\frac{\kappa\varepsilon_+}{2}\,(1+\xi)(1+3\xi)\,
\bigl(r_{\rm tr}^2-r^2\bigr),
\label{eq:sc_p}\\[2pt]
\varepsilon(r)&=\varepsilon_++\frac{p(r)-p_{\rm tr}}{c_{s,+}^2}\,.
\label{eq:app_eps}
\end{align}
The factors $(1+\xi)$ and $(1+3\xi)$ do not go away for a small core,
because the pressure at its boundary is set by the envelope; this is what
distinguishes a quark core from a self-bound strange star, for which
$\xi=0$. The variation of $c_s^2$ inside the core, $(dc_s^2/dp)_+$, would
enter only at second order.

\subsection{Gravitational and baryon masses}

The gravitational mass follows from Eq.~\eqref{eq:Mgr}. Since
$dm_g/dr=4\pi r^2\varepsilon$ contains no metric factor, the only
first-order correction is the mass stored in the density gradient of
Eq.~\eqref{eq:app_eps}; with $\langle r_{\rm tr}^2-r^2\rangle_2=\tfrac25r_{\rm tr}^2$,
\begin{equation}
M_C=\kappa r_{\rm tr}^3\left[1+\frac{(1+\xi)(1+3\xi)}{5\,c_{s,+}^2}\,\kappa r_{\rm tr}^2\right],
\label{eq:app_M}
\end{equation}
which is Eq.~\eqref{eq:sc_M}.

The baryon mass, Eq.~\eqref{eq:MB}, adds the proper-volume factor
$(1-2m_g/r)^{-1/2}\simeq1+\kappa r^2$ and the baryon density $n_B(r)$.
The latter follows from the $T=0$ thermodynamic
identities already used in Eq.~\eqref{eq:deps_dnB},
\begin{equation}
\varepsilon+p=\mu_B\,n_B,\qquad
\frac{dn_B}{dp}=\frac{dn_B}{d\varepsilon}\,\frac{d\varepsilon}{dp}
=\frac{1}{\mu_B\,c_s^2}\,,
\label{eq:app_thermo}
\end{equation}
the second of which uses $d\varepsilon=\mu_B\,dn_B$. At the
interface the first identity gives the quark-side density of
Sec.~\ref{sec:EOS},
$n_B^Q=(\varepsilon_++p_{\rm tr})/\mu_B^{\rm tr}=\varepsilon_+(1+\xi)/\mu_B^{\rm tr}$,
and inside the core
\begin{equation}
\begin{aligned}
n_B(r)&=n_B^Q+\frac{p(r)-p_{\rm tr}}{\mu_B^{\rm tr}\,c_{s,+}^2}\\
&=n_B^Q\left[1+\frac{\kappa}{2\,c_{s,+}^2}\,(1+3\xi)\bigl(r_{\rm tr}^2-r^2\bigr)\right].
\end{aligned}
\label{eq:app_nB}
\end{equation}
The relative gradient of $n_B$ is thus that of $\varepsilon$ divided by
$(1+\xi)$, because $dn_B/n_B=d\varepsilon/(\varepsilon+p)$. Collecting the
leading term, the volume factor
($\langle\kappa r^2\rangle_2=\tfrac35\kappa r_{\rm tr}^2$) and the gradient
term ($\langle r_{\rm tr}^2-r^2\rangle_2=\tfrac25r_{\rm tr}^2$),
\begin{equation}
\begin{aligned}
M_{b,C}={}&\frac{m_N}{\mu_B^{\rm tr}}(1+\xi)\,\kappa r_{\rm tr}^3\\
&\times\left[1+\left(\frac35+\frac{1+3\xi}{5\,c_{s,+}^2}\right)\kappa r_{\rm tr}^2\right].
\end{aligned}
\label{eq:app_Mb}
\end{equation}
Dividing by Eq.~\eqref{eq:app_M} gives Eq.~\eqref{eq:sc_Mb}: the $\tfrac35$ is
the proper-volume excess, and the gradient terms of numerator and
denominator cancel except for the mismatch
$-\xi(1+3\xi)/(5c_{s,+}^2)$ between the growth rates of $n_B$ and
$\varepsilon$.

\subsection{Moment of inertia}

For the moment of inertia, Eq.~\eqref{eq:dI_dr}, the zeroth-order
solution is the uniform Newtonian sphere with the inertial density
$\varepsilon+p$ in place of $\varepsilon$,
\begin{equation}
\mathcal I_0(r)=\tfrac25(1+\xi)\,\kappa r^5 .
\label{eq:app_I0}
\end{equation}
Three first-order corrections multiply the integrand: the profile of the
inertial density, obtained from Eqs.~\eqref{eq:sc_p} and \eqref{eq:app_eps},
\begin{equation}
\varepsilon+p=\varepsilon_+(1+\xi)\left[1+\Bigl(1+\frac1{c_{s,+}^2}\Bigr)
\frac{\kappa}{2}(1+3\xi)\bigl(r_{\rm tr}^2-r^2\bigr)\right],
\label{eq:app_inertial}
\end{equation}
the metric factor, and the frame-dragging factor evaluated with
Eq.~\eqref{eq:app_I0},
\begin{equation}
\Bigl(1-\frac{2m_g}{r}\Bigr)^{-1}\simeq1+2\kappa r^2,\qquad
1-\frac{5\mathcal I_0}{2r^3}\simeq1-(1+\xi)\,\kappa r^2 ,
\label{eq:app_metric}
\end{equation}
the term $\mathcal I^2/r^6$ being of second order. Their
product, linearized, gives
\begin{equation}
\begin{aligned}
\frac{d\mathcal I}{dr}=2\kappa(1+\xi)\,r^4\Bigl[1&+(1-\xi)\,\kappa r^2\\
&+\Bigl(1+\frac1{c_{s,+}^2}\Bigr)\frac{\kappa}{2}(1+3\xi)\bigl(r_{\rm tr}^2-r^2\bigr)\Bigr],
\end{aligned}
\label{eq:app_dIdr}
\end{equation}
where the first correction collects the metric and frame-dragging factors
and the second the profile of the inertial density. Integrating with the
$n=4$ averages of Eq.~\eqref{eq:app_avg},
$\langle r^2\rangle_4=\tfrac57r_{\rm tr}^2$ and
$\langle r_{\rm tr}^2-r^2\rangle_4=\tfrac27r_{\rm tr}^2$,
\begin{equation}
I_C=\frac25(1+\xi)\,\kappa r_{\rm tr}^5\bigl[1+\tilde\beta\,\kappa r_{\rm tr}^2\bigr],\qquad
\tilde\beta=\frac{6-2\xi}{7}+\frac{1+3\xi}{7\,c_{s,+}^2}\,.
\label{eq:app_I}
\end{equation}
Since $\bar I_C\mathcal C_C^2=I_C/(M_C r_{\rm tr}^2)$, the normalization by
Eq.~\eqref{eq:app_M} subtracts $(1+\xi)(1+3\xi)/(5c_{s,+}^2)$ from
$\tilde\beta$ and yields the coefficient $\beta$ of Eq.~\eqref{eq:sc_ab}.

\subsection{Tidal deformability}

The tidal deformability requires the tidal function
$y\equiv rH'/H$, which obeys~\cite{Hinderer:2007mb,Postnikov:2010yn}
\begin{equation}
r\,y'=-y^2-y\,F-r^2Q,
\label{eq:app_ydiff}
\end{equation}
with
\begin{align}
F&=\frac{1-4\pi r^2(\varepsilon-p)}{1-2m_g/r},\notag\\
Q&=\frac{4\pi\bigl[5\varepsilon+9p+(\varepsilon+p)/c_s^2\bigr]-6/r^2}{1-2m_g/r}\notag\\
&\quad-\left[\frac{2\,(m_g+4\pi r^3p)}{r^2\,(1-2m_g/r)}\right]^2,
\label{eq:app_FQ}
\end{align}
and the regular solution starts at $y(0)=2$. Evaluating $F$ and $Q$ to first
order with the zeroth-order profiles (the squared bracket in $Q$ is of second
order),
\begin{equation}
\begin{aligned}
F&=1+(3\xi-1)\,\kappa r^2,\\
r^2Q&=-6+\Bigl[3+27\xi+\frac{3(1+\xi)}{c_{s,+}^2}\Bigr]\kappa r^2,
\end{aligned}
\label{eq:app_FQ1}
\end{equation}
where the last term of $r^2Q$ is the compressibility contribution
$4\pi r^2(\varepsilon+p)/c_s^2$. At zeroth order Eq.~\eqref{eq:app_ydiff}
reads $r\,y'=-(y+3)(y-2)$, whose regular solution is $y=2$: inside a uniform
Newtonian sphere $H\propto r^2$ throughout. Writing $y=2+\delta y$ and
linearizing,
\begin{equation}
r\,\delta y'+5\,\delta y=-\Bigl[1+33\xi+\frac{3(1+\xi)}{c_{s,+}^2}\Bigr]\kappa r^2,
\label{eq:app_dy}
\end{equation}
whose regular solution is $\delta y=-\alpha\,\kappa r^2$, with $\alpha$ as in
Eq.~\eqref{eq:sc_ab} (the homogeneous solution $\propto r^{-5}$ is singular
at the center). At the interface,
\begin{equation}
y_Q(r_{\rm tr})=2-\alpha\,\mathcal C_C+O(\mathcal C_C^2).
\label{eq:sc_y}
\end{equation}

The Love number follows from the closed-form
expression of Refs.~\cite{Hinderer:2007mb,Postnikov:2010yn},
\begin{equation}
k_2(\mathcal C,y)=\frac{8\mathcal C^5}{5}(1-2\mathcal C)^2
\bigl[2-y+2\mathcal C(y-1)\bigr]\,\mathcal D^{-1},
\label{eq:app_k2closed}
\end{equation}
with
\begin{equation}
\begin{aligned}
\mathcal D={}&2\mathcal C\bigl[6-3y+3\mathcal C(5y-8)\bigr]\\
&+4\mathcal C^3\bigl[13-11y+\mathcal C(3y-2)+2\mathcal C^2(1+y)\bigr]\\
&+3(1-2\mathcal C)^2\bigl[2-y+2\mathcal C(y-1)\bigr]\ln(1-2\mathcal C),
\end{aligned}
\label{eq:app_k2den}
\end{equation}
evaluated with $\mathcal C=\mathcal C_C$ and $y=y_Q(r_{\rm tr})$. For any
$y$, the terms of order $\mathcal C$ to $\mathcal C^4$ of $\mathcal D$
cancel identically and its leading term is $\tfrac{16}{5}(3+y)\,\mathcal C^5$,
so that $k_2\to(2-y)/[2(3+y)]$, the Newtonian Love number, as
$\mathcal C\to0$. For the core, $y\to2$ as $\mathcal C_C\to0$ and the
leading term of the numerator vanishes as well: $k_2^C$ is of first order
in the compactness. Expanding numerator and denominator one order further
with $y=2-\alpha\,\mathcal C_C$ one finds
\begin{equation}
k_2^C=\Bigl(\frac15+\frac{\alpha}{10}\Bigr)\mathcal C_C+O(\mathcal C_C^2),
\label{eq:app_k2}
\end{equation}
and $\Lambda_C\mathcal C_C^4=\tfrac23\,k_2^C/\mathcal C_C$ gives
Eq.~\eqref{eq:sc_L}. The structure of Eq.~\eqref{eq:app_k2} is transparent in
the Newtonian limit: the term $\alpha\,\mathcal C_C/10$ is the Newtonian
response $(2-y)/[2(3+y)]$ to the departure of $y$ from 2, and
$\mathcal C_C/5$ is the first post-Newtonian correction, present even for
$y=2$. The same limit gives $k_2=\tfrac34$ for $y=-1$, the value reached if
the surface correction of a density drop to vacuum, which lowers $y$ by 3
at small compactness [Eq.~\eqref{eq:junction_y} with
$\Delta\varepsilon=\varepsilon_+$ and $p_{\rm tr}=0$], were applied at
$r_{\rm tr}$. The first-order coefficient of $\Lambda_C\mathcal C_C^4$
requires $y$ to second order and is not given here; the empirical function
$f(\mathcal C_C)$ of Eq.~\eqref{eq:sc_f} takes its place.

%==============================================================

\bibliography{references}

\end{document}